\documentclass[]{aastex631}
\usepackage{graphicx}
\usepackage{enumitem}
\usepackage{footnote}
\setlist[itemize]{noitemsep}

\begin{document}

\title{Characterization of blue and yellow straggler stars of Berkeley 39 using \textit{Swift}/UVOT}

\author[0009-0008-0887-6636]{Komal Chand}
\thanks{p20210463@pilani.bits-pilani.ac.in} 
\affiliation{Department of physics, Birla Institute of Technology and Science-Pilani, 333031 Rajasthan, India.}

\author[0000-0001-7470-9192]{Khushboo Rao}
\thanks{khushboo@gm.astro.ncu.edu.tw} 
\affiliation{Department of physics, Birla Institute of Technology and Science-Pilani, 333031 Rajasthan, India.}
\affiliation{Institute of Astronomy, National Central University, 300 Zhongda Road,  Zhongli 32001 Taoyuan, Taiwan.}

\author{Kaushar Vaidya}
 \thanks{kaushar@pilani.bits-pilani.ac.in} 
\affiliation{Department of physics, Birla Institute of Technology and Science-Pilani, 333031 Rajasthan, India.}

\author{Anju Panthi}
\thanks{p20190413@pilani.bits-pilani.ac.in} 
\affiliation{Department of physics, Birla Institute of Technology and Science-Pilani, 333031 Rajasthan, India.}



\begin{abstract}

We characterize blue straggler stars (BSS) and yellow straggler stars (YSS) of an open cluster (OC) Berkeley 39 using multi-wavelength observations including \textit{Swift}/UVOT. Our analysis also makes use of ultraviolet (UV) data from \textit{GALEX}, optical data from \textit{Gaia} DR3 and Pan-STARRS, and infrared data from 2MASS, \textit{Spitzer}/IRAC, and \textit{WISE}. Berkeley 39 is a $\sim$6 Gyr old Galactic OC located at a distance of $\sim$4200 pc. We identify 729 sources as cluster members utilizing a machine learning algorithm, ML-MOC, on \textit{Gaia} DR3 data. Of these, 17 sources are classified as BSS candidates and four as YSS candidates. We construct multi-wavelength spectral energy distributions (SEDs) of 16 BSS and 2 YSS candidates, within the \textit{Swift}/UVOT field, to analyze their properties. Out of these, 8 BSS candidates and both the YSS candidates are successfully fitted with single-component SEDs. Five BSS candidates show marginal excess in the near-UV (fractional residual $<$ 0.3 in 
all but one UVOT filter), whereas three BSS candidates show moderate to significant excess in the near-UV (fractional residual $>$ 0.3 in at least two UVOT filters). We present the properties of the BSS and YSS candidates, estimated based on the SED fits.

\end{abstract}



\section{Introduction} \label{sec:intro}

Binary stars, ranging from close binaries to wide binaries, are common in star clusters. They play a significant role in star clusters, contributing to their dynamical and evolutionary processes. The interactions among close binary stars and multiple stellar systems can lead to intriguing, unique stellar populations in star clusters. Notable outcomes of such interactions are blue straggler stars (BSS, \citealt{Sandage1953}), yellow straggler stars (YSS, \citealt{Strom1971, Leiner2016}), red straggler stars \citep{Geller2017}, cataclysmic variables \citep{Ritter2010}, etc. BSS are located above the main sequence turn-off (MSTO) in the clusters' colour-magnitude diagrams (CMDs). This location on the CMDs indicates that they have higher temperatures and luminosities than the MSTO. BSS are observed in diverse stellar environments such as open clusters (OCs, \citealt{ahumada2007new, Jadhav2021, Rain}), globular clusters (GCs, \citealt{Sandage1953, Nikitha}), Galactic fields \citep{carney2005metal} and dwarf galaxies \citep{momany2007blue}. Observational studies have shown that BSS are among the most massive members of star clusters \citep{shara1997first, gilliland1998oscillating, fiorentino2014blue}. Therefore, they are supposed to segregate towards the clusters' centers faster than any other cluster population. The radial distributions of BSS are used to categorize GCs and OCs in different dynamical ages \citep{ferraro2012dynamical, Vaidya2020, Rao2021, Rao2023}. Thus, BSS provide valuable insights to probe the relationship between stellar evolution and stellar dynamics \citep{bailyn1995blue}. Although much rarer compared to the BSS, YSS are frequently found in CMDs as objects relatively redder than the BSS and brighter than the subgiants \citep{Leiner2016}. The first YSS was discovered in an OC M67 by \cite{Janes}. \cite{Landsman1997} studied this object, a yellow giant, S1040, using the Hubble Space Telescope, and found a white dwarf (WD) companion associated with it. Recently, \cite{Rain} have identified 77 potential YSS among 408 OCs using the \textit{Gaia} DR2 data \citep{brown2018gaia}. Hot companions of few YSS have been indentified in OCs, NGC 2506 \citep{Panthi2022}, NGC 6940  \citep{Anju6940} and NGC 2627 \citep{Saketh_2024}  using the \textit {Astrosat}/UVIT  \citep{Astrosat} data.\\

Formation mechanisms of BSS are still debatable and are an active field of research \citep{Boffin2015}. There are three widely accepted formation pathways of BSS, direct stellar collisions, mass transfer (MT) in a binary, and mergers. Direct stellar collisions may lead to formation of a single massive BSS \citep{hills1976stellar, Leonard1996}. In hierarchical triple systems, components of the inner binary may have a MT or merger as a result of the Kozai mechanism, leading to the formation of a binary BSS or single massive BSS, respectively \citep{perets2009triple}. MT efficiency can be conservative or non-conservative, depending on the orbital periods and mass ratio of a binary system \citep{leiner2021census}. Observational evidence of BSS with WD companions in low-density environments such as OCs raised the possibility of MT as a possible pathway \citep{G}. Based on the evolutionary phase of the primary star, the MT process is further divided into three cases. In Case A, a compact binary system consisting of two main-sequence (MS) stars may undergo merger via stellar winds or a MT process, leaving behind a single massive BSS, or it may undergo MT via Roche lobe overflow, leaving behind a BSS with an MS companion \citep{Vilhu}. In Case B, the MT occurs in a long-period binary system (P $\sim$ 1 -- 1000 days) when the primary is in the RGB phase, leaving behind a binary BSS with a He WD companion \cite{Landsman1997}. Finally, in Case C, the MT occurs in a wide binary system (P $\sim$ 1000 -- 2000 days) when the primary is in the AGB phase, leaving behind a binary BSS with a Carbon-Oxygen (CO) WD companion \citep{Gosnell}. There is a fourth possible pathway of BSS formation via MT. In wide binary systems with periods ranging from $\sim$ 1000 -- 10000 days, MT may also occur via wind Roche lobe overflow (WRLOF, \citealt{Geller}) when the primary is in the AGB phase (Case D), leaving behind a binary BSS with a CO WD companion. As mentioned above, various mechanisms contribute to BSS formation, and determining the precise pathway is challenging without details about their orbital parameters and companions. Multiple mechanisms have been found to be dominant in a given cluster \citep{Ferraro, M}. In low-density environments such as OCs and Galactic fields, MT and merger pathways are dominant, while high-density environments such as the cores of GCs may also form BSS via the collision pathway \citep{Davies}. YSS are considered the evolved BSS, i.e., the post MS stars more massive than the MSTO, on their way to becoming the RGB stars \citep{Mathieu1990}. Through an asteroseismic mass and radius measurements of a YSS S1237 in OC M67, \cite{Leiner2016} concluded that it has formed either due to stellar collision or binary merger. \\

Ultraviolet (UV) wavelengths are more suitable for studying exotic populations like BSS and YSS. Being hotter objects, they emit a significant fraction of their flux in the UV wavelengths. Moreover, binary BSS and YSS with a WD companion would show an excess flux in the UV, which may be used to characterize them. Stellar hot spots in contact and semi-detached binaries also enhance UV flux \citep{kouzuma2019starspots}. Magnetic activity and chromospheric processes also contribute to the total UV flux emitted by single and binary stars of intermediate-age star clusters. Additionally, flares on the stars serve as a transient source of UV radiation and in X-rays \citep{dempsey1993rosat, mitra2005relationship}. Hence, it is important to understand the source of UV excess flux in known binary systems, which could be due to the intrinsic property of the star or due to the presence of a hot companion. \cite{G} detected 4 hot WD companions and 3 cool WD companions (11,000 K -- 12,000 K) for BSS of NGC 188 using Hubble Space Telescope (HST) observations in far-ultraviolet (FUV) filters. Similar work has been done for other OCs using \textit {Astrosat/UVIT} observations \citep{Subramaniam2016,subramaniam2018ultraviolet,Jadhav2021,Panthi2022,Vaidya7789, Anju2024} as well as \textit{Swift/UVOT} observations \citep{Rao2022}. Recently, \cite{Anju6940} also studied WD companions for other exotic populations, including YSS, blue lurkers (BLs), and red clump stars of NGC 6940 using \textit {Astrosat/UVIT} data.\\

Berkeley 39 ($\alpha$\ = $7^h46^m42^s$, $\delta$= $-4^h36^m00^s$ ) is a 6 Gyr \citep{Kassis1997} old OC having mass $\sim 2\times 10^{4} M_{\odot}$ and mean metallicity [Fe/H] = $-$0.20 \citep{Bragaglia2012}, located at a distance of 4200 pc \citep{Vaidya2020}. The BSS population of this cluster has been previously studied by \cite{Vaidya2020} and \cite{Rao2023} using the \textit{Gaia} EDR3 data \citep{Gaia2021}. \cite{Vaidya2020} identified 23 BSS whereas \cite{Rao2023} identified 17 BSS. They found this cluster to be dynamically young or intermediate age based on the segregation of its BSS with respect to other less massive reference populations. This cluster has been subjected to various photometric and spectroscopic studies \citep{Bragaglia2012}, along with the identification of BSS. However, it has never been studied using UV imaging observations. Characterization of two BSS of this cluster using multi-wavelength SEDs, including the Swift/UVOT data, are presented in \cite{Chand}. With an aim to characterize the BSS and YSS populations of this cluster using multi-wavelength spectral energy distributions (SEDs), we study this cluster.\\
The paper is arranged as follows. In §\ref{sec:2}, we give information about the data used for the present work and membership determination of the cluster. In §\ref{sec:3}, we present the results obtained. Finally, in §\ref{sec:4}, we present the discussion and summary of the work.

 \section{Data and Membership Identification}\label{sec:2}

In order to characterize the BSS and YSS populations, we use multiwavelength data, ranging from UV to infrared (IR). In the subsections below, we provide the details of the archival data from various surveys and the cluster membership determination using the machine learning-based membership algorithm, ML-MOC \citep{Agarwal2021}, on the \textit{Gaia} DR3 data \citep{GaiaDR3}.
\subsection{UV data}
The fluxes of sources in UV wavelengths are obtained from the Ultraviolet and Optical Telescope (\textit{Swift}/UVOT; \citealt{gehrels2004swift}) and Galaxy Evolution Explorer (\textit{GALEX}; \citealt{Martin2005}). \textit{Swift} spacecraft is equipped with three main instruments, namely Burst Alert Telescope (BAT), sensitive in the energy range of 3.0 -- 150.0 keV, X-ray Telescope (XRT), sensitive in the energy range of 0.3 -- 10.0 keV, and UVOT, sensitive in the optical/UV band in the wavelength range of 170 -- 650 nm. UVOT is a modified Ritchey-Chretien UV/optical telescope having a diameter of 30 cm and provides a wide-field view of $17^{\prime} \times 17^{\prime}$. This instrument utilizes various filters, including clear white filters, U (300 -- 400 nm), B (380 -- 500 nm), V (500 -- 600 nm), UVW1 (220 -- 400 nm), UVM2 (200 -- 280 nm), and UVW2 (180 -- 260 nm), two grisms, a magnifier and a blocked filter \citep{roming2005swift}. We utilize near-UV (NUV) data \citep{Siegel2019} from UVOT in this work. Berkeley 39 was observed by \textit{Swift}/UVOT in three UV filters, UVW2, UVM2, and UVW1, in 2011. The details of observations, the total number of detections, and the ML-MOC (described in Section \ref{sec:2.4}) counterparts of cluster members in all three filters are listed in Table \ref{Table1}. In addition to the \textit{Swift}/UVOT data, we also use the \textit{GALEX} data to study our targets. \textit{GALEX} is a 50 cm aperture space-based telescope \citep{Martin2005} that operates in two UV bands, FUV (1350 -- 1780 Å) and NUV (1770 -- 2730 Å). For Berkeley 39, fluxes are available in the NUV filter alone.

\begin{table}[htbp]
	\centering
	\caption{Observation details of OC Berkeley 39 using \textit{Swift}/UVOT}
	\begin{tabular}{ccccc}
		\hline
		Filter & $\lambda_{eff}$ & Exposure & Detections & Counterparts\\
		& (\AA) & (seconds) &  & (ML-MOC members)\\ 
		\hline
		UVW2 & 2085.73 & 522 & 178 & 40  \\
		UVM2 & 2245.78 & 637 & 243 & 35  \\
		UVW1 & 2684.14 & 592 & 519 & 166\\
		\hline
	\end{tabular}
	\label{Table1}
\end{table}

\subsection{Optical Data}
\textit{Gaia} data \citep{GaiaDR3} provides proper motions, parallaxes, and three-band photometry (G, G$_{BP}$, G$_{RP}$) for over a billion sources. The wavelength of filter G is centred at 673 nm, G$_{BP}$ at 532 nm, and G$_{RP}$ at 797 nm \citep{2010A&A...523A..48J}. The Panoramic Survey Telescope and Rapid Response System (Pan-STARRS; \citealt{schlafly2012photometric}) is a ground-based telescope. It utilises five filters, g, r, i, z, and y, with effective wavelengths 481, 617, 752, 866, and 962 nm \citep{tonry2012pan}, respectively. 

\subsection{IR Data}
Two Micron All Sky Survey (2MASS; \citealt{Cohen2003}) contains near-IR photometric measurements in the J (1.24 $\mu$m), H (1.66 $\mu$m), and $K_s$ (2.16 $\mu$m) bands. \textit{Spitzer}/IRAC is a space-based observatory containing a four-channel camera that provides mid-IR wavelength fluxes \citep{fazio2004infrared}. IRAC provides $5.^{\prime}2 \times 5.^{\prime}$2 simultaneous broadband images at I1 (3.6 $\mu$m), I2 (4.5 $\mu$m), I3 (5.8 $\mu$m), and I4 (8.0 $\mu$m). In 2006, \textit{spitzer}/IRAC observed Berkeley 39 in all four IRAC channels with an exposure time of 26.8 s. \cite{fazio2004infrared} provides information about sources detected in all four filters, I1, I2, I3, and I4. Wide-field Infrared Survey Explorer (\textit{WISE}; \citealt{Wright2010}) is a mid-IR full sky survey. It includes a 40 cm diameter telescope and four infrared detectors, namely, W1, W2, W3, and W4, with wavelengths centred at 3.35 $\mu$m, 4.60 $\mu$m, 11.56 $\mu$m, and 22.09 $\mu$m, respectively. Since the \textit{Spitzer}/IRAC filters have better resolution than WISE, we use WISE W1, W2 fluxes only when \textit{Spitzer}/IRAC fiuxes are unavailable.


\begin{figure*}
	\begin{minipage}{0.3\textwidth}
		\includegraphics[width=\textwidth]{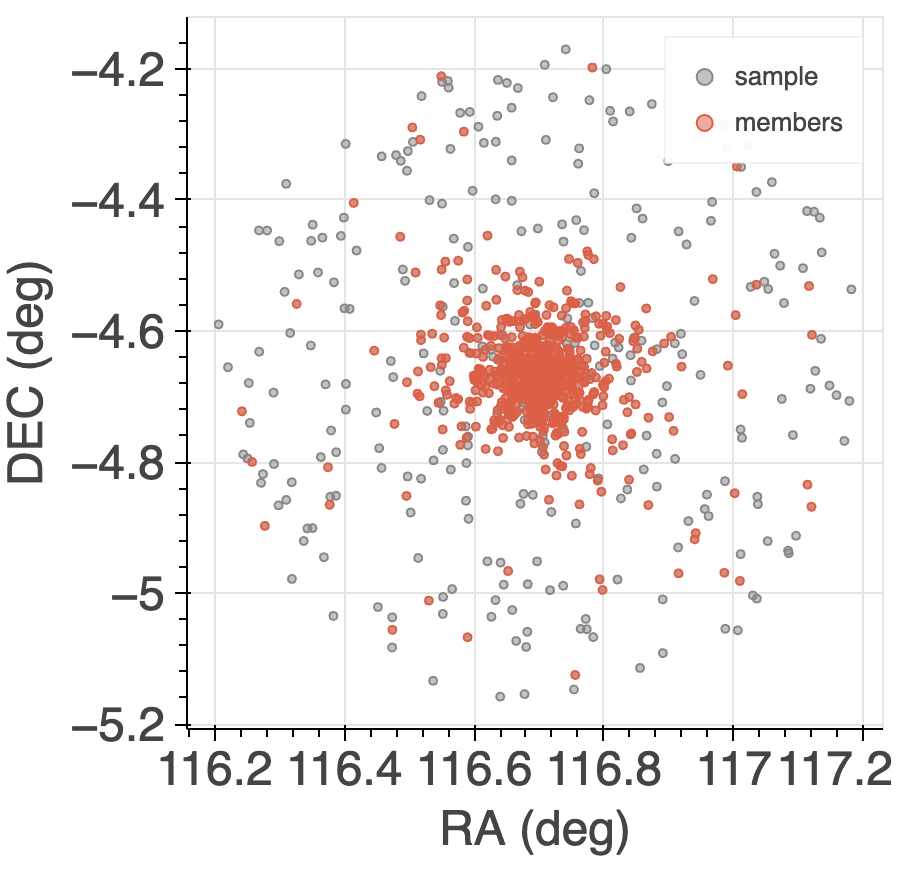}
		\centering
		(a)
		
	\end{minipage}%
	\begin{minipage}{0.3\textwidth}
		\includegraphics[width=\textwidth]{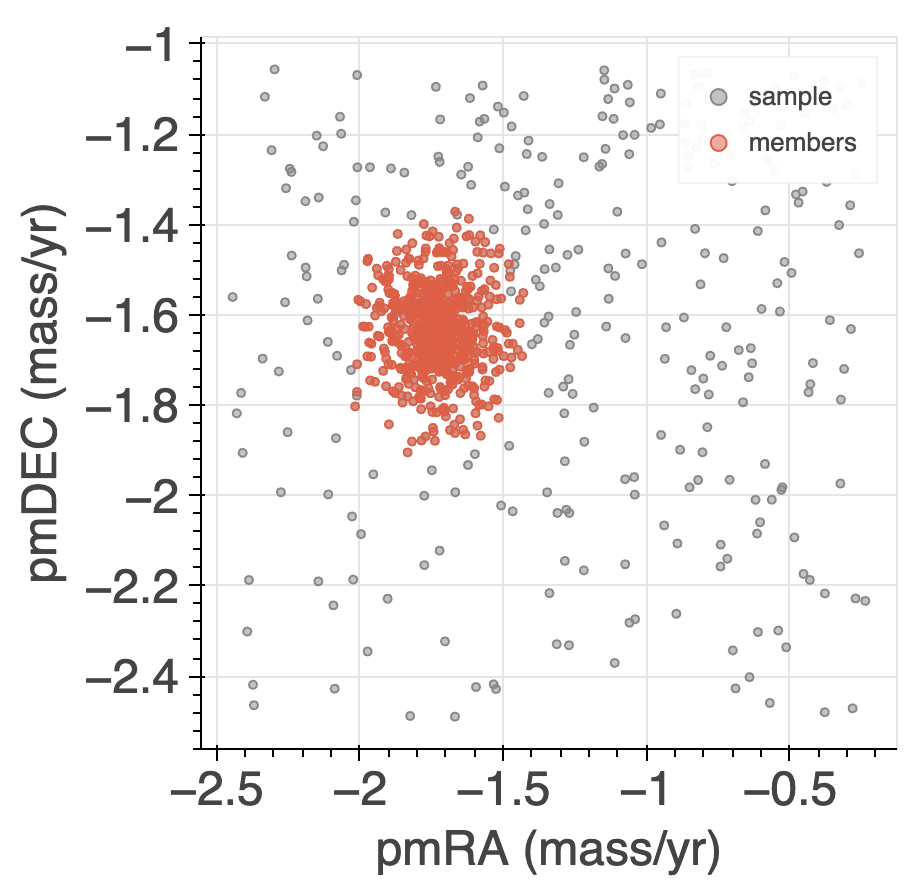}
		\centering
		(b)
	\end{minipage}%
	\begin{minipage}{0.3\textwidth}
		\includegraphics[width=\textwidth]{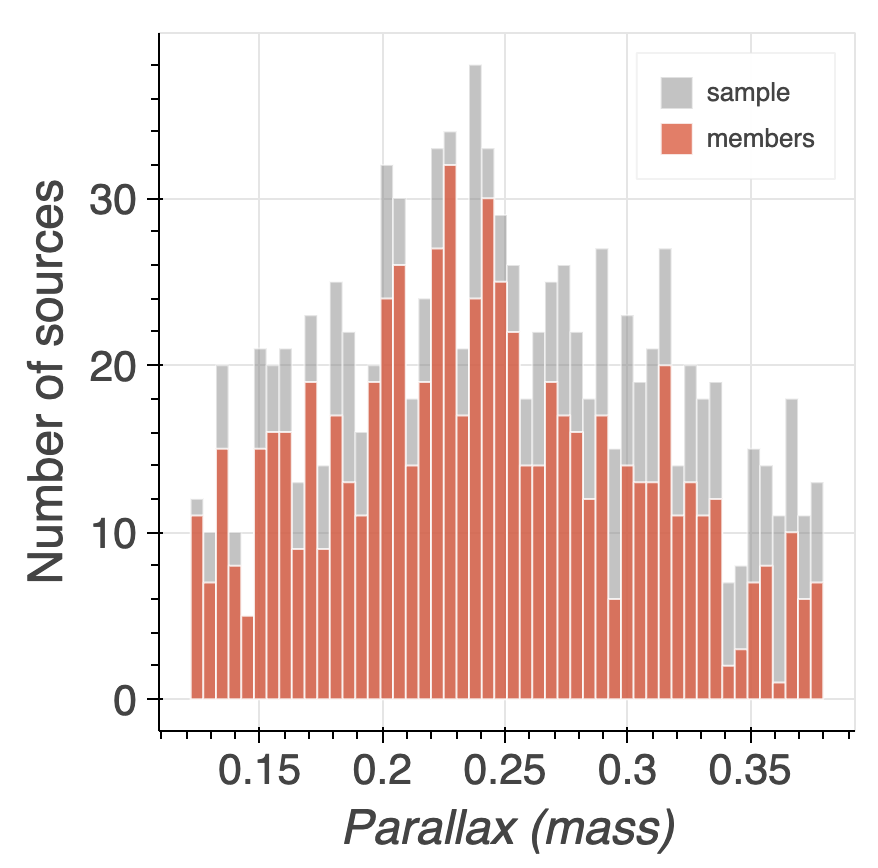}
		\centering
		(c)
	\end{minipage}%
	\hfill
	\caption{(a) The spatial distribution (b) proper motion distribution and (c) parallax distribution using ML-MOC algorithm on \textit{Gaia} DR3 data.}
	\label{Figure1}
\end{figure*}

\subsection{Membership Identification}\label{sec:2.4}
We use a machine-learning based membership determination algorithm, ML-MOC, for determination of cluster members  \citep{Agarwal2021}. ML-MOC uses parallax and proper motion from \textit{Gaia} data to identify cluster members and does not require prior knowledge about the cluster. It is based on the k-Nearest Neighbour (kNN, \citealt{Cover1967}) and the Gaussian mixture model (GMM, \citealt{Peel2000}) algorithms.
The steps followed to determine the cluster members are summarised as follows. Initially, we download sources within 30$^{\prime}$ radius from the cluster center. These sources are required to have five astrometric parameters (RA, DEC, pmRA, pmDEC, and parallax), appropriate measurements in the \textit{Gaia} photometric passbands (G, G$_{BP}$, and G$_{RP}$), non-negative parallaxes, and G-magnitude errors less than 0.005 mag. This group of sources is denoted as \textit{All sources}. Subsequently, we employ kNN algorithm to eliminate probable field stars and determine \textit{Sample sources} with a higher fraction of cluster members. Then, we apply the GMM to the \textit{Sample sources} to separate the likely cluster members from the field stars. The GMM also provides the membership probabilities for each source. First, we select high-probability members with membership probabilities greater than 0.6, and then we include sources with membership probabilities between 0.2 and 0.6 to obtain the complete catalog of the cluster. For further details on the ML-MOC algorithm, readers are referred to \cite{Agarwal2021}. We obtained 659 sources with cluster membership probabilities $\geq$ 0.6 and additional 70 sources with probabilities between 0.2 and 0.6. By plotting the radial distribution of identified cluster members, we estimated the cluster radius to be 14$^{\prime}$. The cluster radius estimated by us is in agreement with that estimated by \cite{Rao2023}. Figure \ref{Figure1} shows the spatial, proper motion, and parallax distribution of sample sources and cluster members of Berkeley 39.

\begin{figure*}[htbp]
	\centering
	\begin{minipage}{0.49\textwidth}
		\centering
		\includegraphics[width=\textwidth]{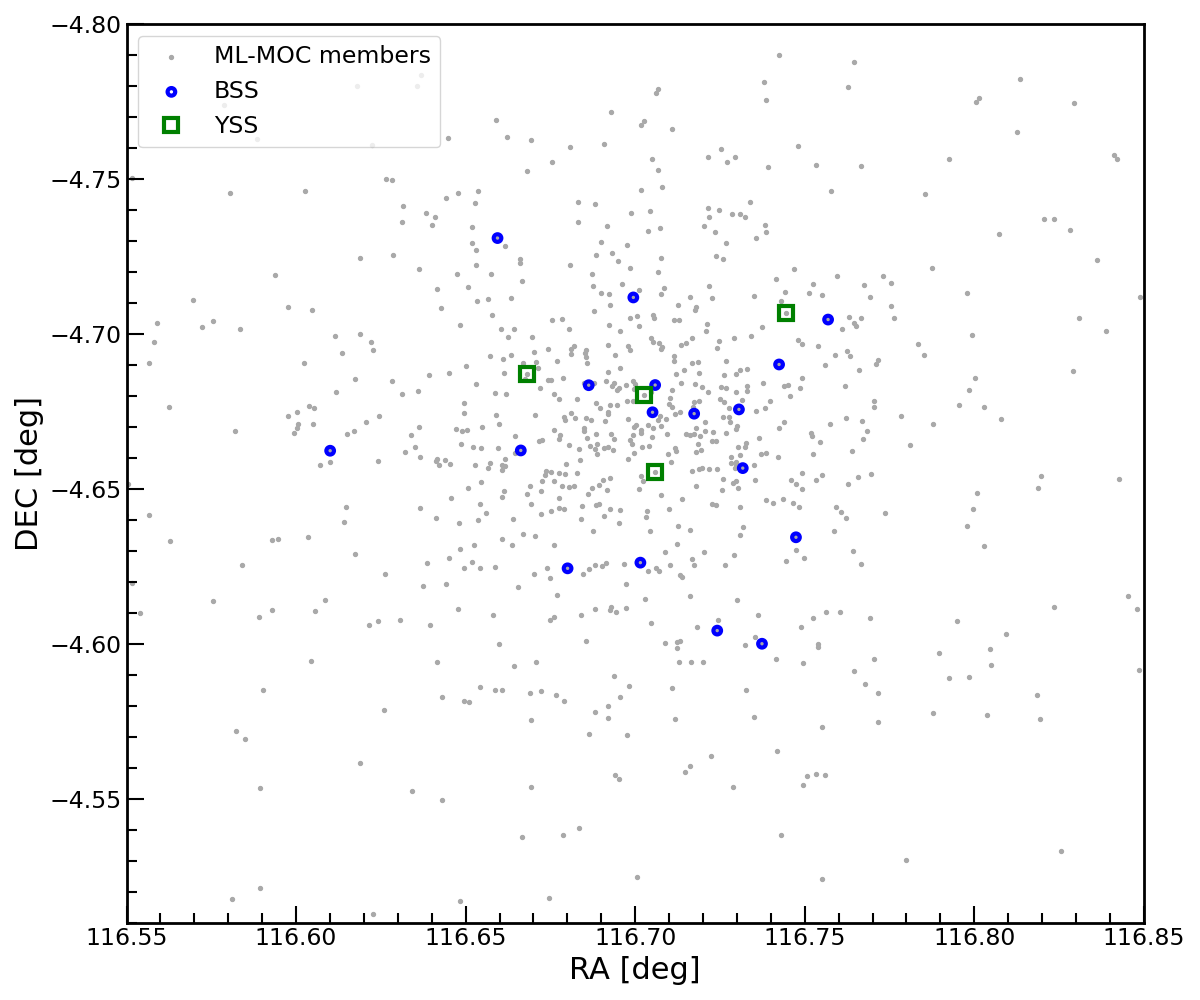}
		(a)
		
	\end{minipage}%
	\hfill
	\begin{minipage}{0.49\textwidth}
		\centering
		\includegraphics[width=\textwidth]{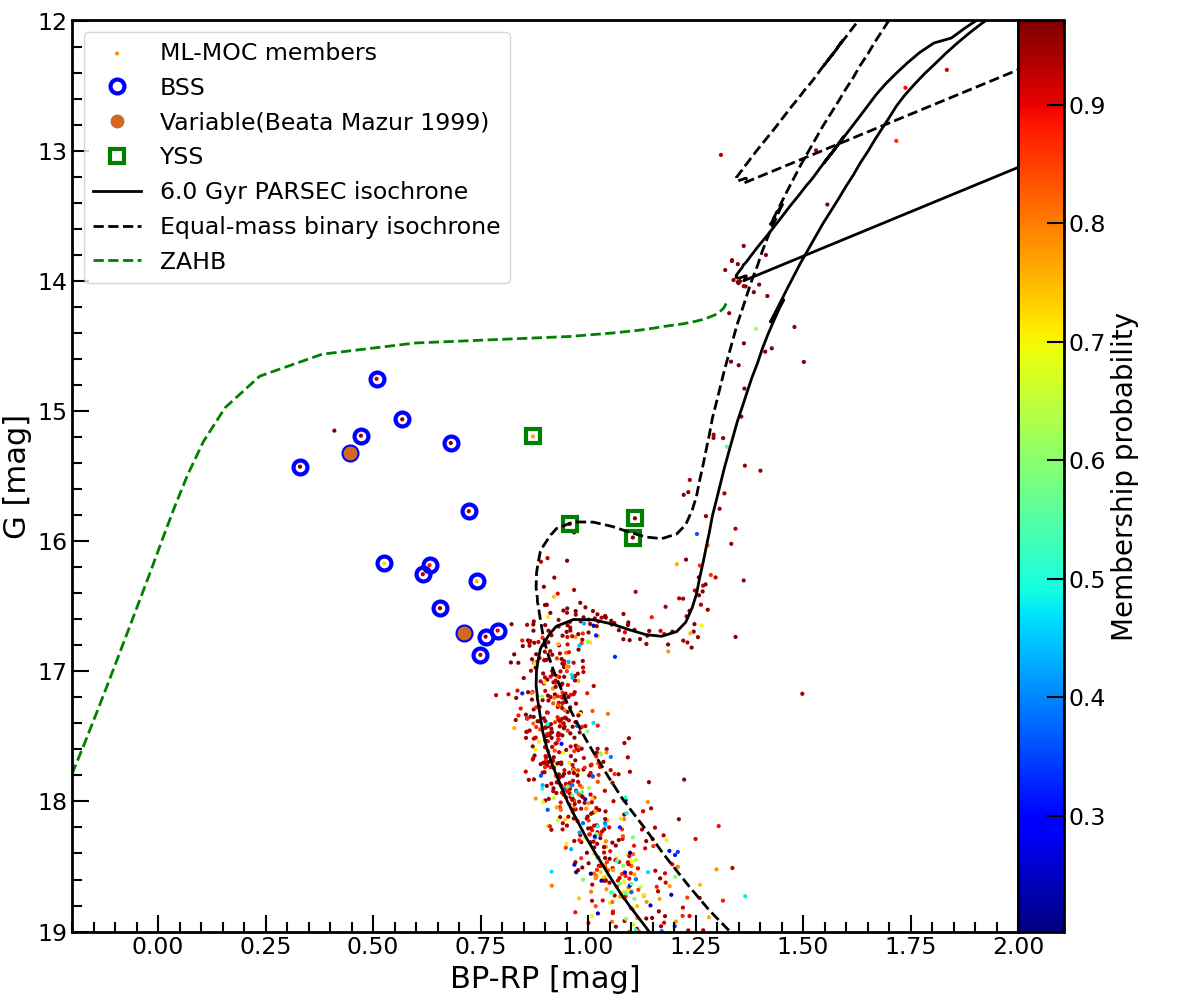}
		(b)
	\end{minipage}%
	\hfill
	\caption{(a) Spatial distribution of the cluster members as grey data points, BSS marked with blue open circles and YSS marked with green open squares. (b) Optical CMD of cluster members of Berkeley 39 identified using the ML-MOC algorithm on \textit{Gaia} DR3 members. The cluster members, along with their probabilities, are shown with auxiliary colours. The BSS and YSS are represented by blue open circles and green open squares, respectively. A PARSEC isochrone \citep{bressan2012parsec} of age = 6 Gyr, distance = 4200 pc, extinction, $A_{\mathrm{G}}$  = 0.41 mag, and color-excess E(BP$-$RP) = 0.16 mag is overplotted on the CMD. The ZAHB is shown as a green dashed line and black dashed line shows equal-mass binary isochrone.}
	\label{Figure2}
\end{figure*}

\section{RESULTS}\label{sec:3}
\subsection{ The Colour Magnitude Diagram}\label{sec:3.1}
\label{A}
Figure \ref{Figure2}a shows the spatial distribution of cluster members within 14$^{\prime}$ radius, with the BSS and YSS marked on the figure. The optical CMD for cluster members obtained using ML-MOC algorithm on \textit{Gaia} DR3 data is shown in Figure \ref{Figure2}b. Additionally, a PARSEC isochrone \citep{bressan2012parsec} is overlaid on the CMD using the known age, 6 Gyr \citep{Kassis1997} and metallicity, [Fe/H] = $-$0.20 from \cite{Bragaglia2012}. 
For distance, extinction, $A_{\mathrm{G}}$, and reddening, E(BP$-$RP), mean distance for bright (G $<$ 16 mag) members using \cite{Bailer2021}, and median values of the $A_{\mathrm{G}}$ and E(BP$-$RP) were used. To fit the isochrone, some fine-tuning of parameters was necessary. The final parameters of the isochrone are an age of 6 Gyr, distance of 4200 parsec, extinction, $A_{\mathrm{G}}$, of 0.41 mag, and color-excess E(BP$-$RP), of 0.16 mag. The fundamental parameters used here to plot isochrone are consistent with those in  \cite{Rao2023}. Among the 729 cluster members, 17 stars are identified as BSS candidates. Additionally, 4 cluster members are found to lie between the region occupied by the BSS and the RGB. These are identified as the YSS candidates. The magnitudes of the BSS candidates range from 0 -- 2 magnitudes brighter as compared to the MSTO. As seen in the Figure \ref{Figure2}, BSS candidates appear in two separate groups (15.2 mag $<$ G $<$ 16.8 mag) and (16.2 mag $<$ G $<$ 17 mag). However, all BSS candidates are seen well below the horizontal branch, and hence, we consider them to be the BSS candidates.

Out of these 17 BSS candidates and 4 YSS candidates, one BSS candidate and one YSS candidate lie outside the field of view of \textit{Swift/UVOT}. For one YSS candidate, UV data is available only in one UVOT filter. Hence, we include 16 BSS candidates and 2 YSS candidates in our analysis. Of these, 5 BSS candidates and one YSS candidate are detected in three filters of \textit{Spitzer}/IRAC (I1, I2, and I3), 10 BSS candidates are detected in at least two IRAC bands, and one BSS candidate and one YSS candidate are not detected in any of the \textit{Spitzer}/IRAC filters. We use the data from WISE W1 and W2 bands for objects when Spitzer/IRAC data are not available. Furthermore two BSS, BSS6 and BSS10, are identified as variables by \cite{Beata}, shown as brown filled circles in the CMD. BSS6 has been identified as a possible eclipsing binary, whereas BSS10 has been identified as a $\delta$-Scuti variable. \textit{Gaia} DR3 lists them as variables; however, their variable nature is not identified. We searched for their TESS data, but they are not observed via TESS.

\begin{figure*}
	\centering
	\begin{minipage}{0.49\textwidth}
		\centering
		\includegraphics[width=\textwidth]{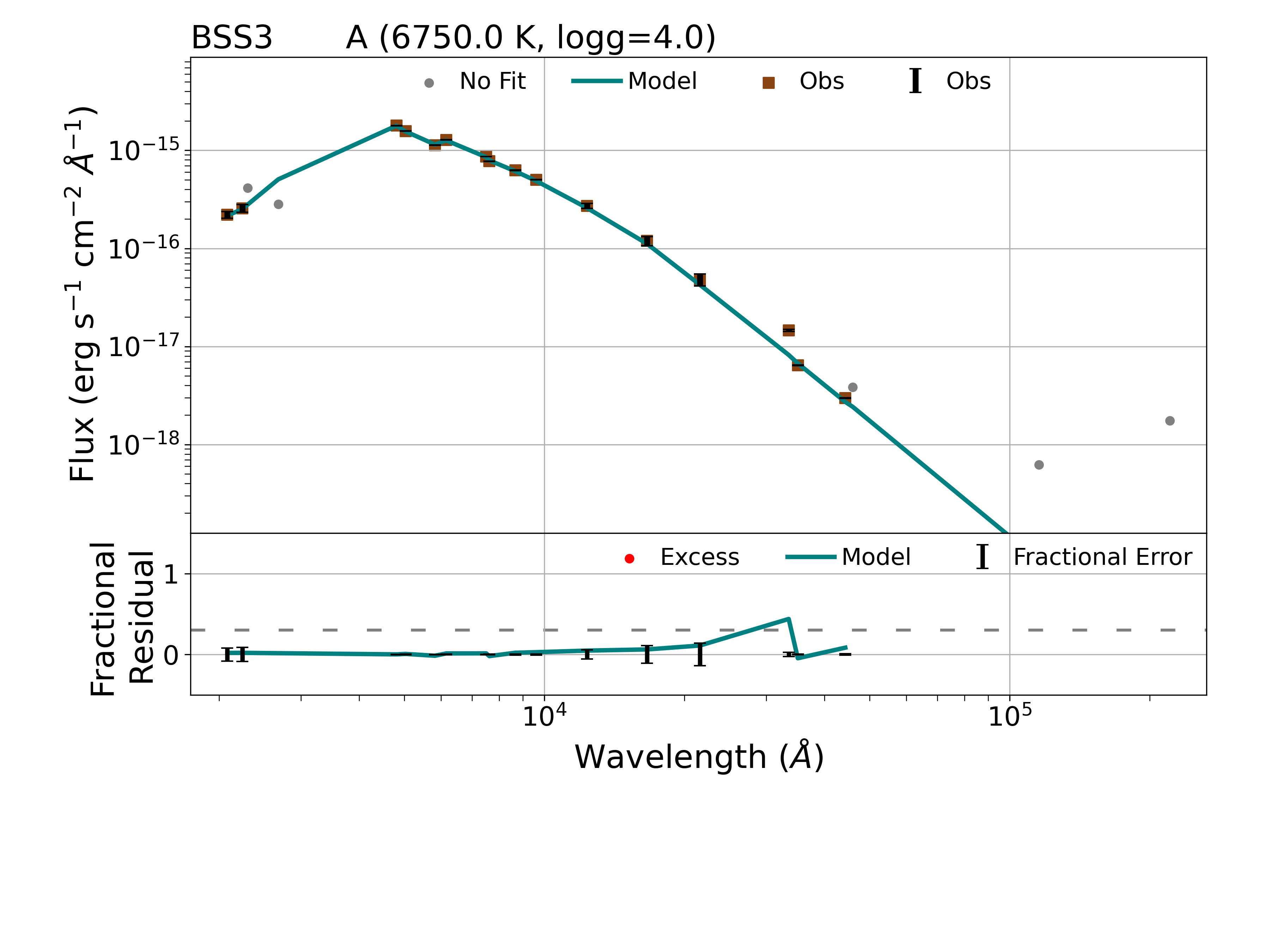}
		
	\end{minipage}%
	\hfill
	\begin{minipage}{0.49\textwidth}
		\centering
		\includegraphics[width=\textwidth]{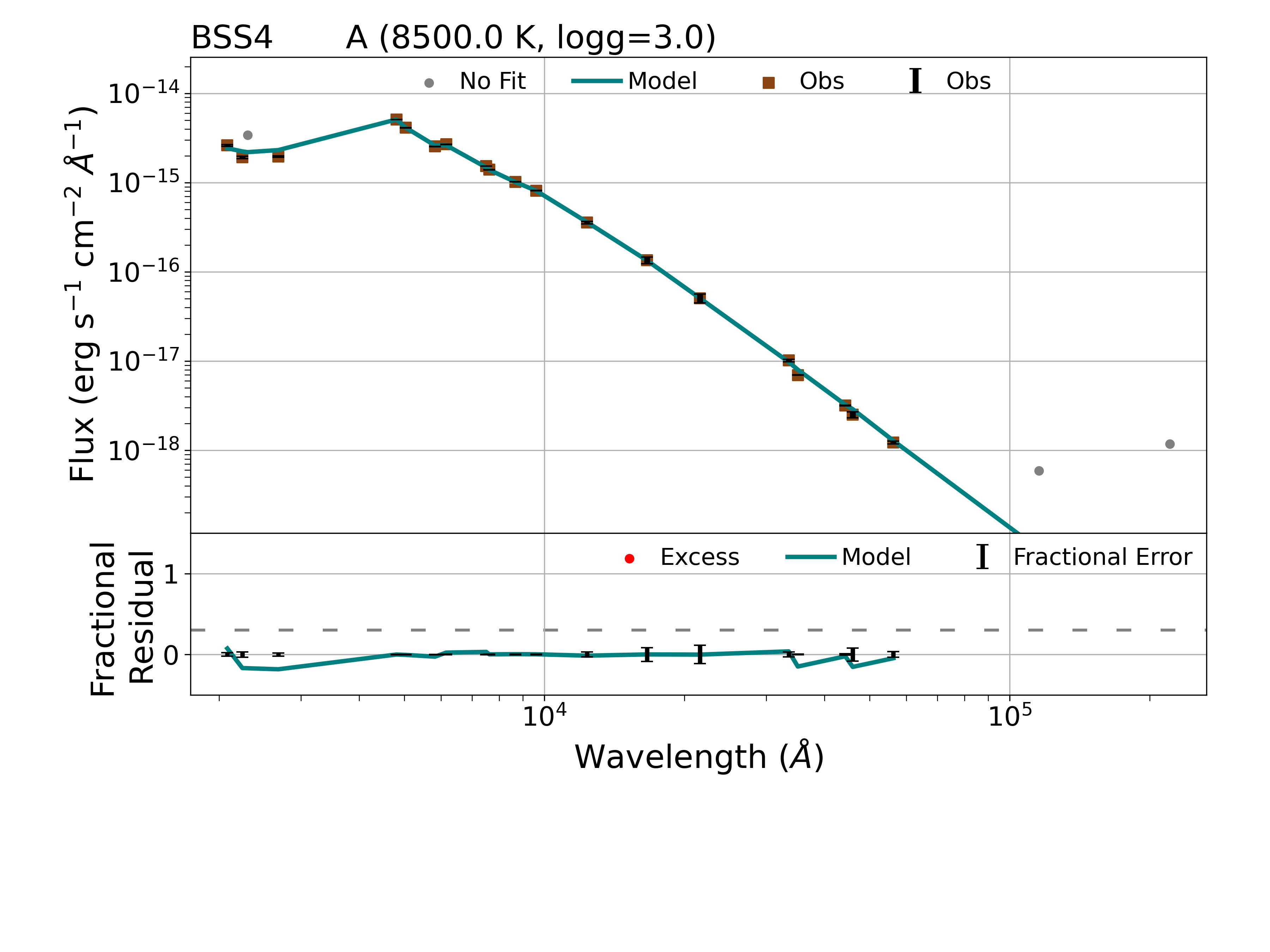}
		
	\end{minipage}%
	\hfill
	\begin{minipage}{0.49\textwidth}
		\centering
		\includegraphics[width=\textwidth]{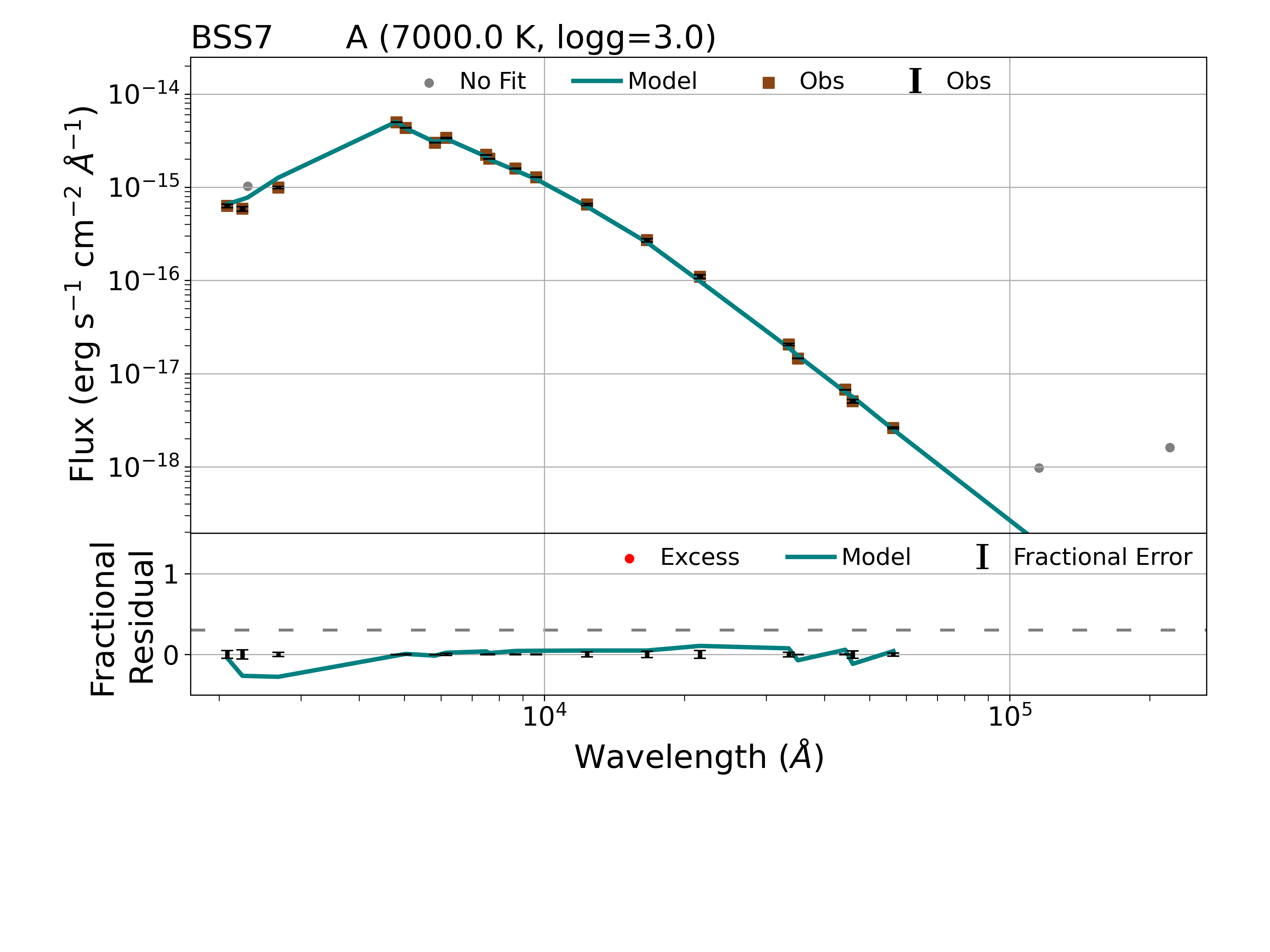}
		
	\end{minipage}%
	\hfill
	\begin{minipage}{0.49\textwidth}
		\centering
		\includegraphics[width=\textwidth]{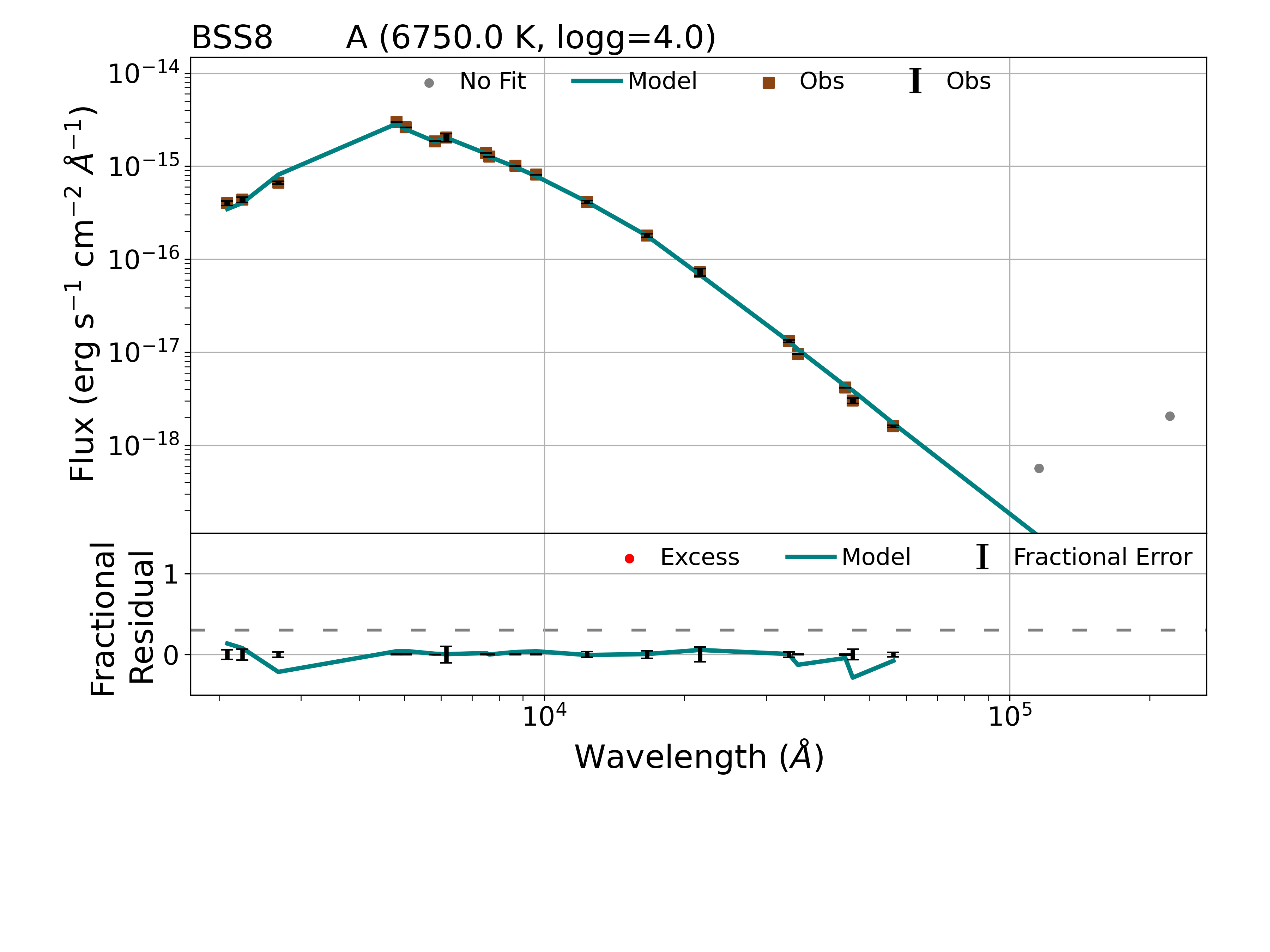}
		
	\end{minipage}%
	\hfill
	\begin{minipage}{0.49\textwidth}
		\centering
		\includegraphics[width=\textwidth]{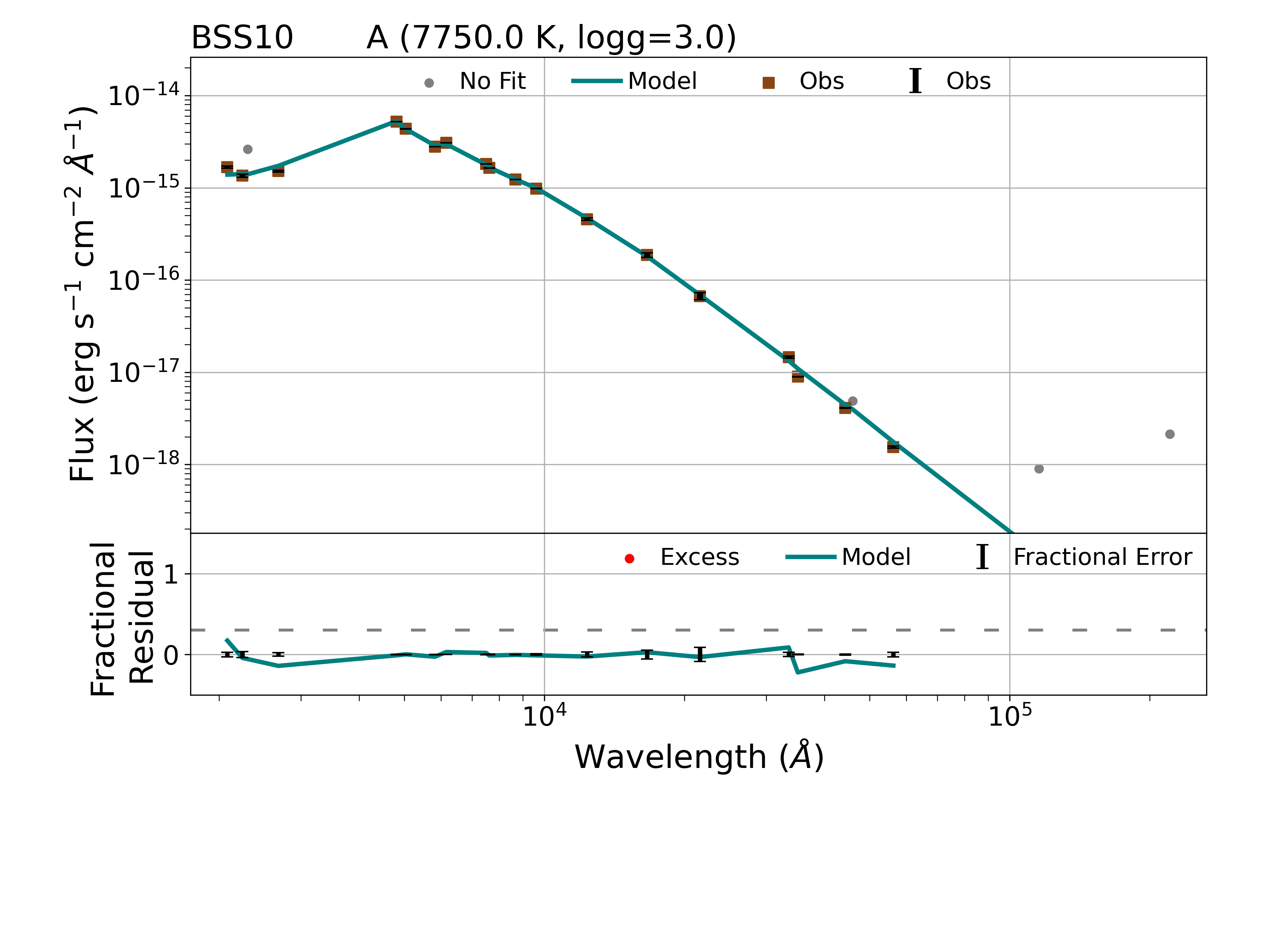}
		
	\end{minipage}%
	\hfill
	\begin{minipage}{0.49\textwidth}
		\centering
		\includegraphics[width=\textwidth]{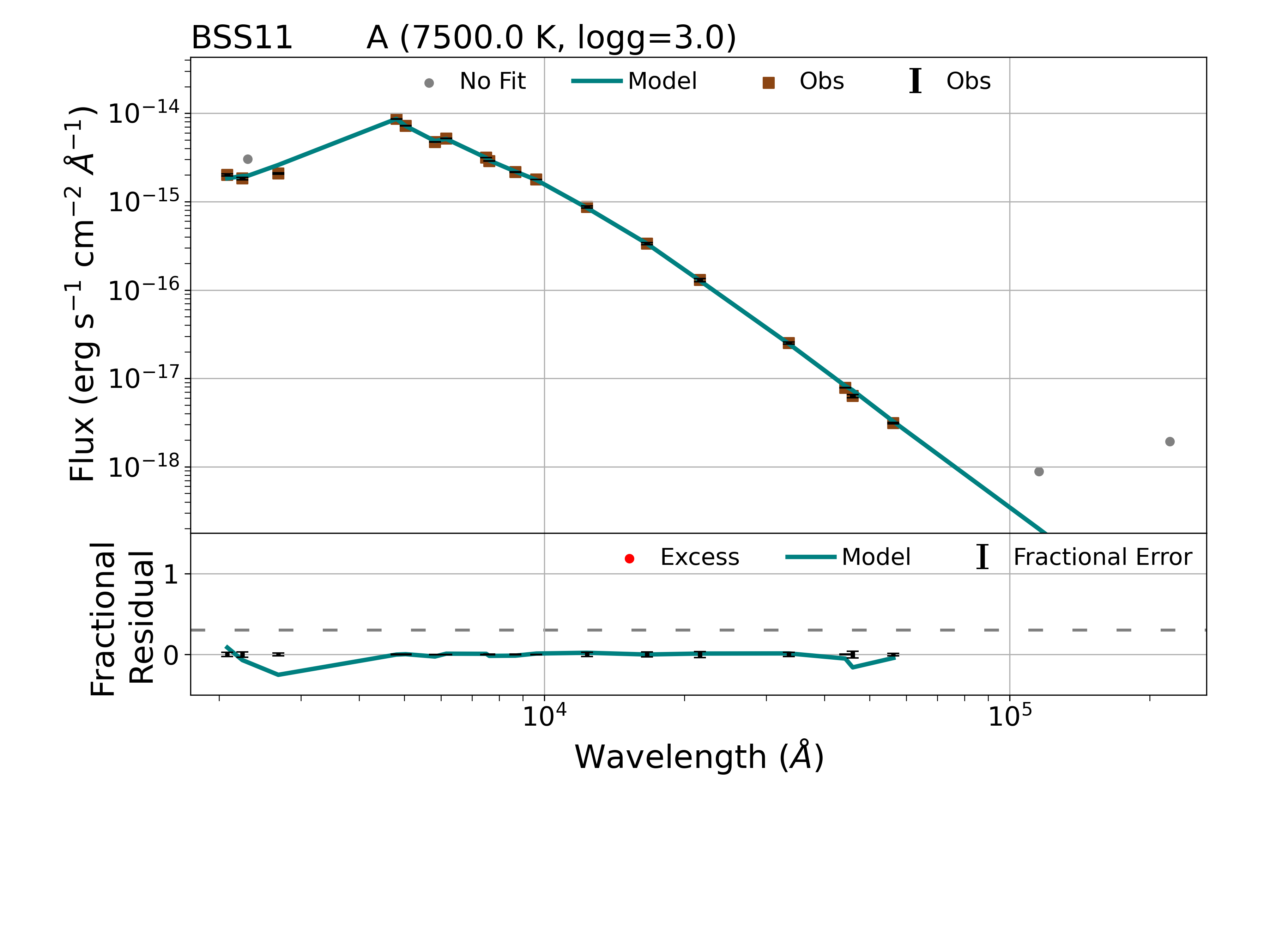}
		
	\end{minipage}%
	\hfill
	\begin{minipage}{0.49\textwidth}
		\centering
		\includegraphics[width=\textwidth]{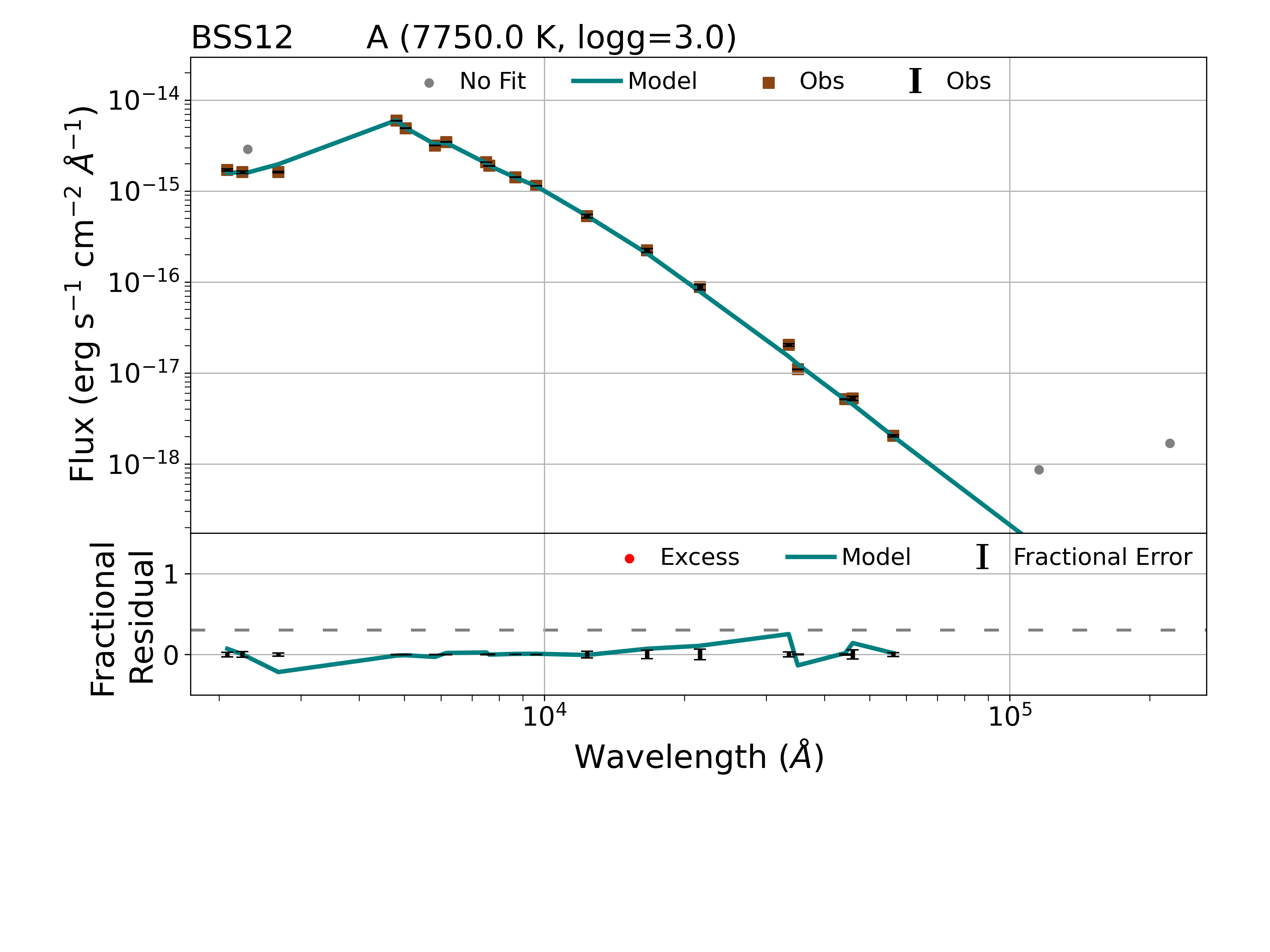}
		
	\end{minipage}%
	\hfill
	\begin{minipage}{0.49\textwidth}
		\centering
		\includegraphics[width=\textwidth]{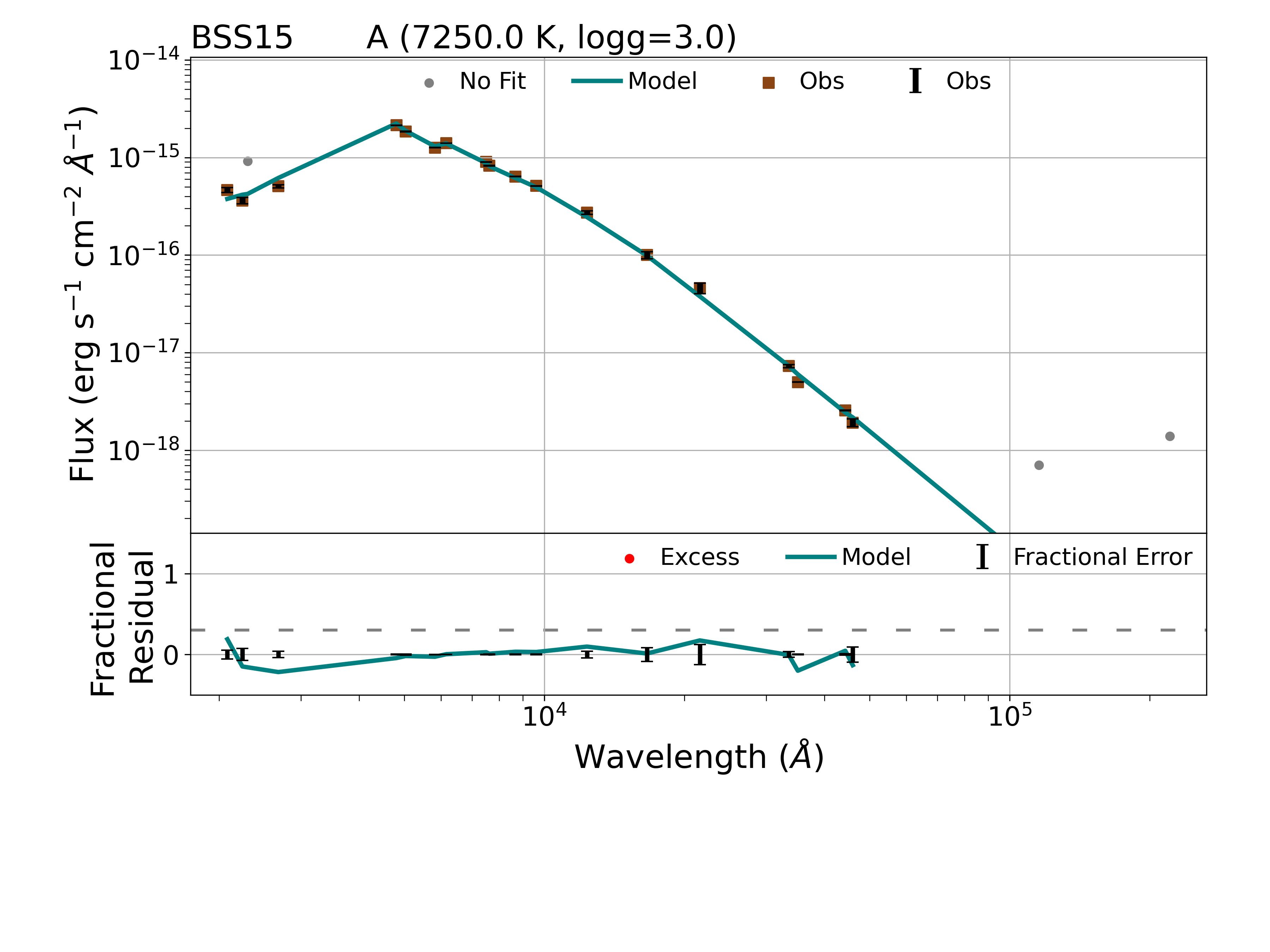}
		
	\end{minipage}%
		\hfill
\end{figure*}

\begin{figure*}
	\centering	
	\begin{minipage}{0.49\textwidth}
		\centering
		\includegraphics[width=\textwidth]{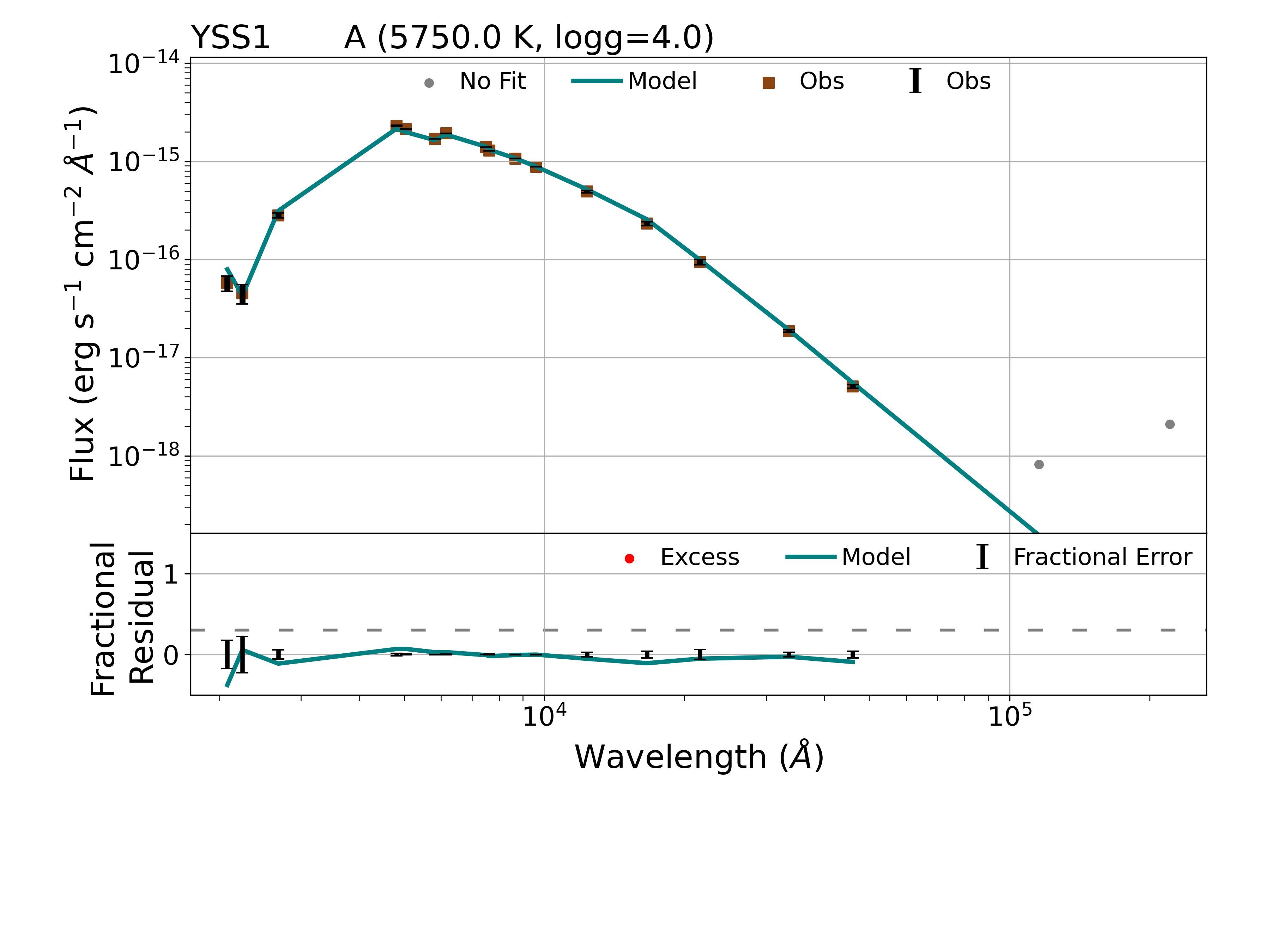}
		
	\end{minipage}%
	\hfill
	\begin{minipage}{0.49\textwidth}
		\centering
		\includegraphics[width=\textwidth]{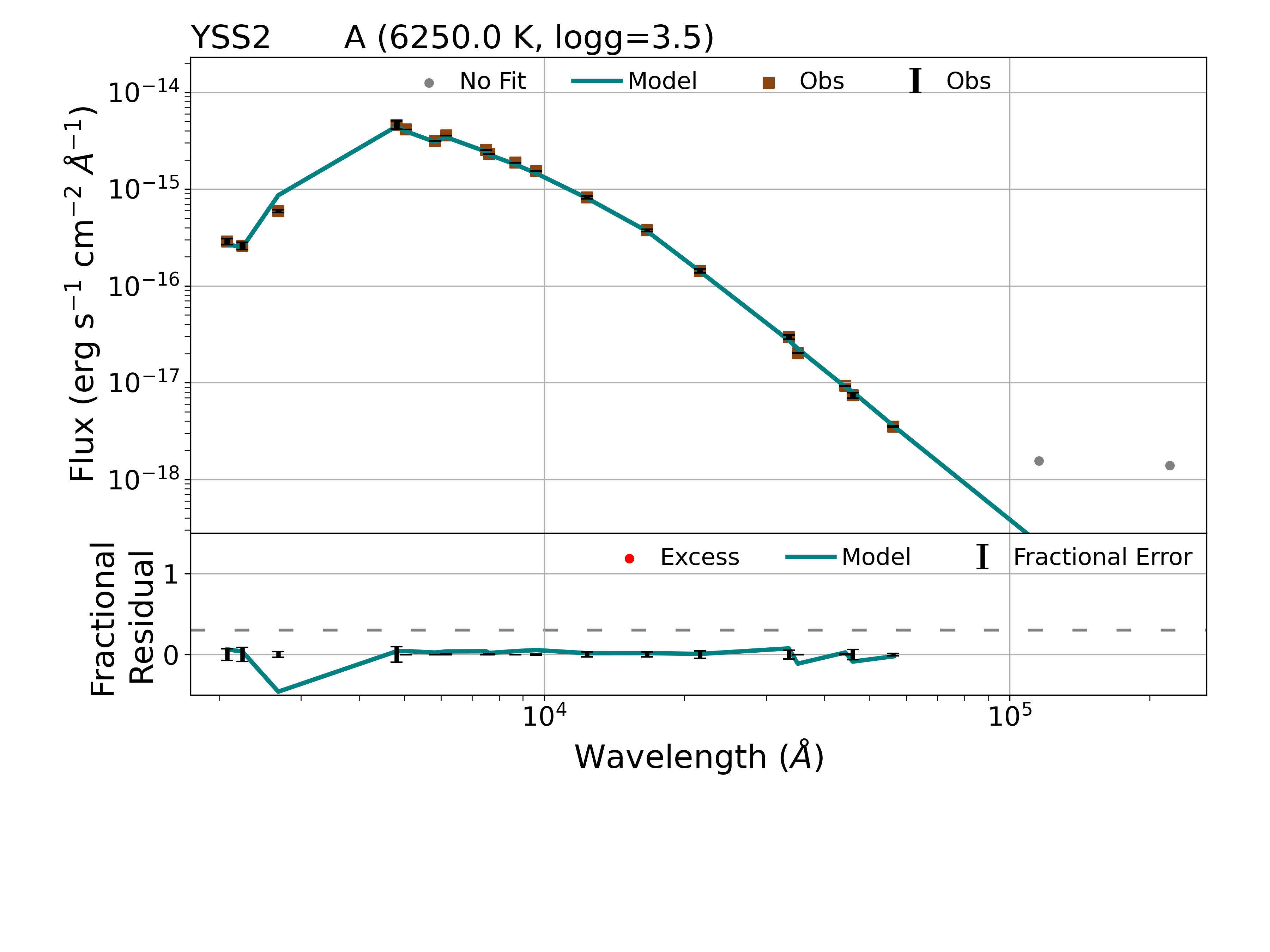}
		
	\end{minipage}%
	\hfill
	\caption{The single-component SED fits of BSS and YSS candidates. The top panels in each figure show extinction corrected observed fluxes in brown data points, with black error bars representing the errors in observed fluxes, and the blue curve representing the Kurucz stellar model fit. The bottom panels depict the residual between extinction-corrected observed fluxes and the model fluxes across the filters from UV to IR wavelengths.The grey data points are not included in the fitting. The dashed line in the lower panel represents a fractional residual of 0.3.}
	\label{Figure3}
\end{figure*}

\subsection{Spectral Energy Distributions}\label{sec:3.2}

\begin{figure*}
	\centering
	\begin{minipage}{0.45\textwidth}
		\centering
		\includegraphics[width=\textwidth]{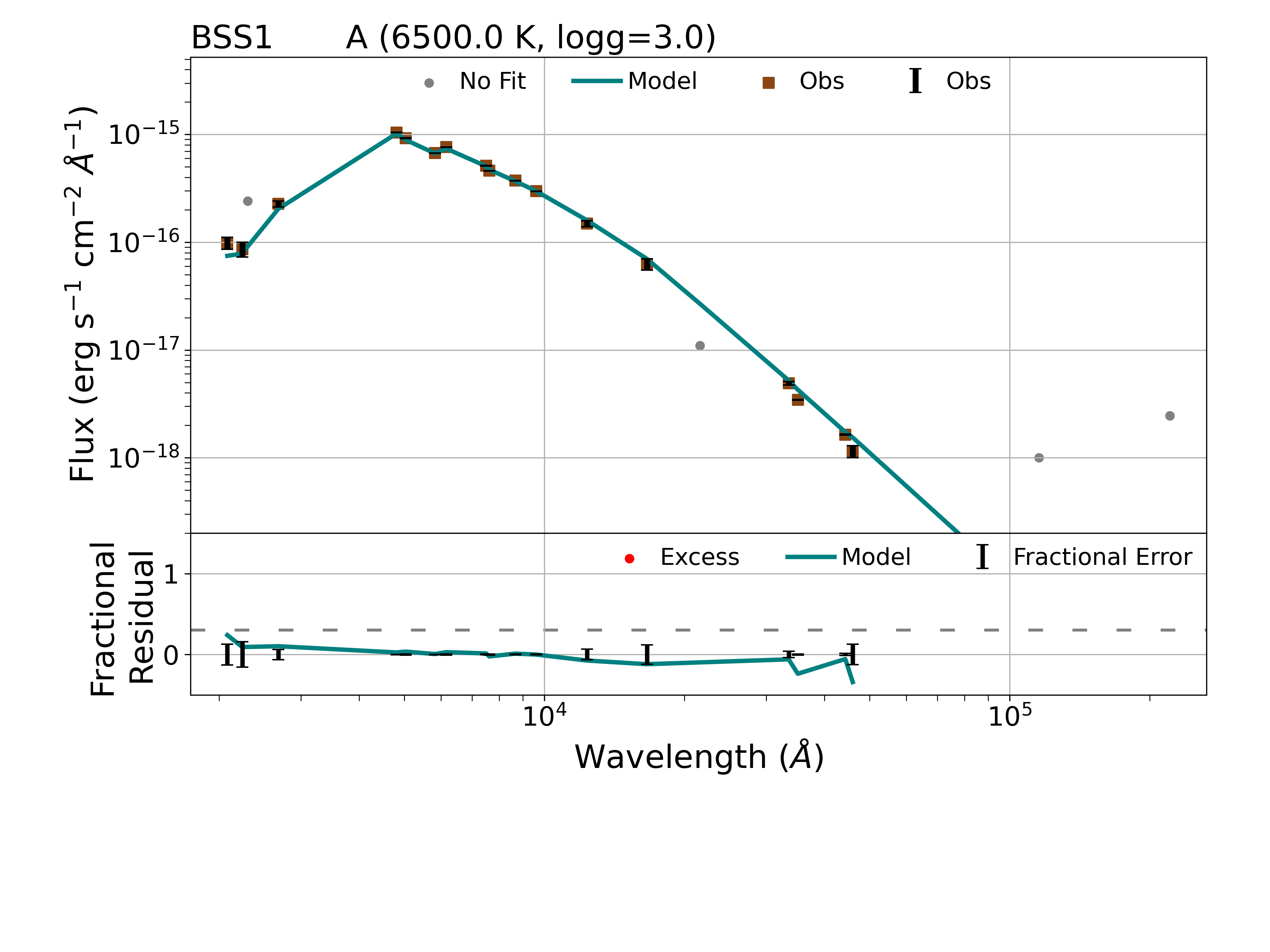}
		
	\end{minipage}%
	\hfill
	\begin{minipage}{0.45\textwidth}
		\centering
		\includegraphics[width=\textwidth]{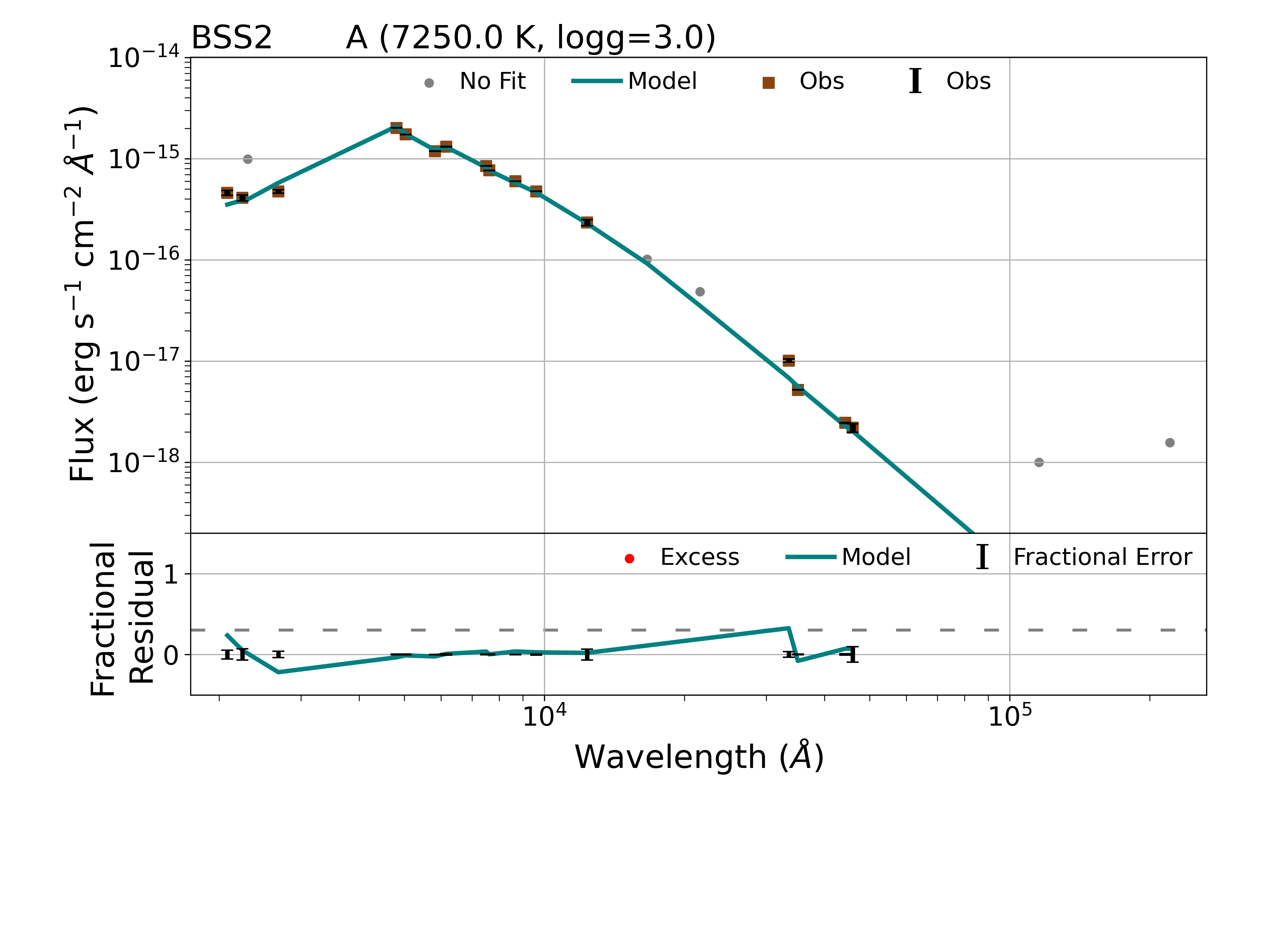}
		
	\end{minipage}%
	\hfill
	\begin{minipage}{0.45\textwidth}
		\centering
		\includegraphics[width=\textwidth]{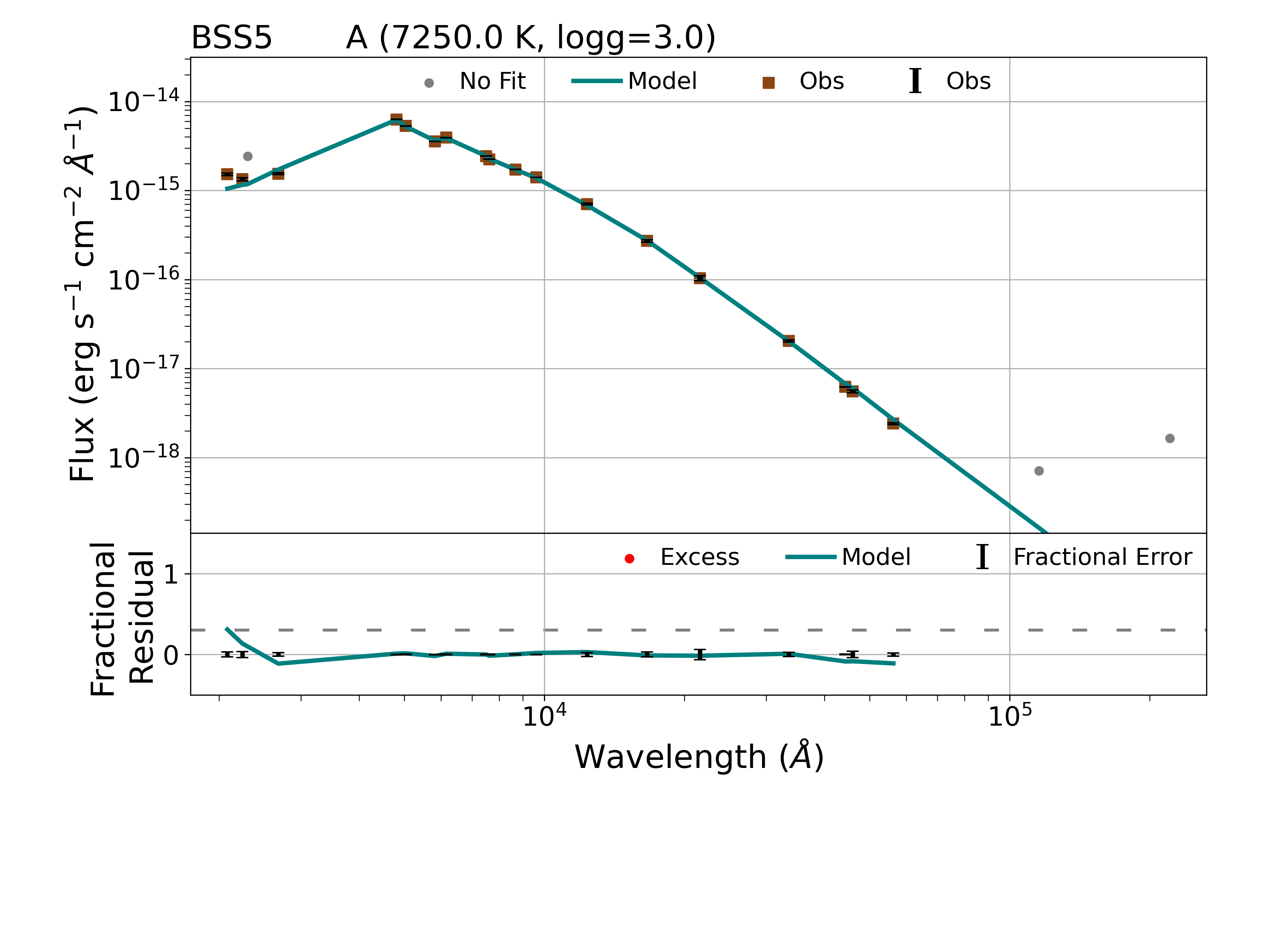}
		
	\end{minipage}%
	\hfill
	\begin{minipage}{0.45\textwidth}
		\centering
		\includegraphics[width=\textwidth]{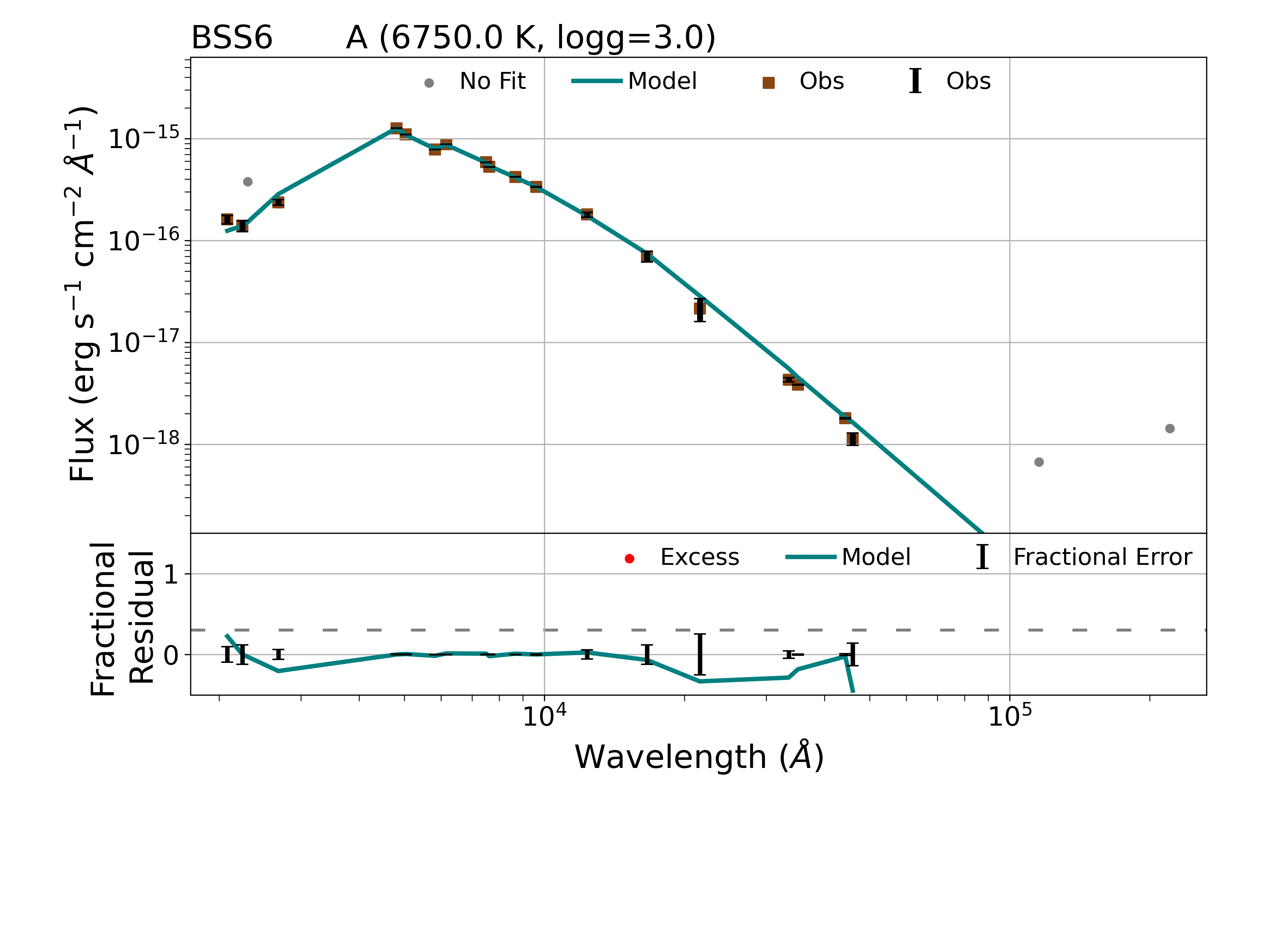}
		
	\end{minipage}%
	\hfill
	\begin{minipage}{0.45\textwidth}
		\centering
		\includegraphics[width=\textwidth]{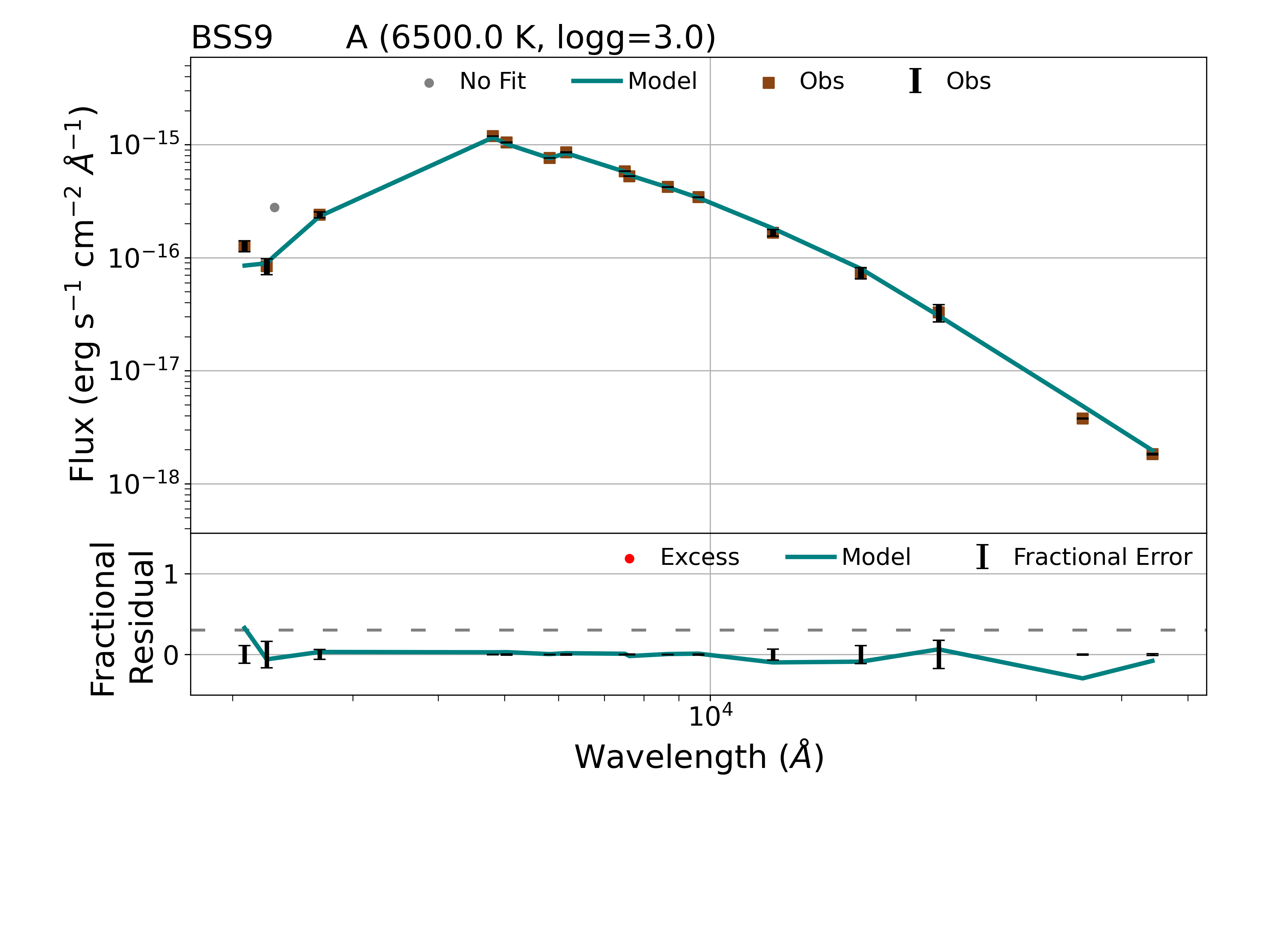}
		
	\end{minipage}%
	\hfill
	\caption{The single-component SED fits of BSS having fractional residual less than 0.3 in all but in one UVOT filter. The symbols and curves mean the same as in Figure  \ref{Figure3}.}
	\label{Figure4}
\end{figure*}

Multi-wavelength SEDs have been a vital tool in characterizing exotic stellar populations such as BSS and YSS. To estimate the fundamental parameters such as temperatures, radii, and luminosities of BSS and YSS candidates, we construct their multi-wavelength SEDs. Before we fit the SEDs, we utilize Aladin\href{https://aladin.cds.unistra.fr/}{\textsuperscript{1}}, a visualization tool, to examine any nearby sources within a 3$^{\prime \prime}$ radial distance. None of the 16 BSS and 2 YSS candidates have a nearby source within 3$^{\prime \prime}$; therefore, we construct their SEDs. The SEDs are constructed using the Virtual Observatory SED Analyzer (VOSA\href{http://svo2.cab.inta-csic.es/theory/vosa/}){\textsuperscript{2}}. VOSA employs filter transmission curves to calculate synthetic photometry for a selected theoretical model. To generate SEDs, we provide distance to the cluster, 4200 pc, and extinction, $A_{\mathrm{V}}$, 0.51 mag. The extinction, $A_{\mathrm{V}}$, was derived from the median $A_{\mathrm{G}}$ of the cluster, using the correlation given in \cite{Wang}. VOSA collects fluxes in optical wavelength from \textit{Gaia} DR3 and Pan-STARRS, near IR from 2MASS, and mid-IR data from \textit{WISE}. We use the \textit{WISE} W1 and W2 band fluxes for sources that do not have IRAC magnitudes in the I1 and I2 channels since the resolution of WISE is poorer compared to that of the IRAC. When only upper limits are available in any filters, they are not included in the fitting of the SEDs. VOSA corrects all of the observed fluxes for extinction using the extinction law given by \cite{Fitzpatrick} and \cite{Indebetouw}. To fit SEDs, we use Kurucz \citep{Castelli1997} stellar model. To ensure the reliability of SED fitting and to check the true flux of sources at UV wavelengths, we first exclude UV data points and fit the SEDs from optical to IR wavelength. VOSA suggests the best-fit SEDs based on the reduced $\chi^2$ minimization. The value of $\chi^2$ is calculated using the following formula:
\begin{equation}
	\chi_r^2=\frac{1}{N-N_f}\sum_{i=1}^{N}\frac{(F_{o,i}-M_dF_{m,i})^2}{\sigma^{2}_{o,i}}\\
\end{equation}
Where $N$ is the total number of photometric data points, $N_f$ indicates the total number of fitted model parameters, $F_{o,i}$ is the observed flux, $F_{m,i}$ is the model flux of the star, $\sigma_{o,i}$ is the error in observed flux, and $M_d=\left(R/D\right)^{2}$ is the scaling factor, dependent on the radius and the distance of objects. The spatial coordinates, \textit{Gaia} IDs, and \textit{Swift}/UVOT fluxes, along with their errors, are listed in Table \ref{Table2}. \\

\begin{figure*}
	\centering
	\begin{minipage}{0.49\textwidth}
		\centering
		\includegraphics[width=\textwidth]{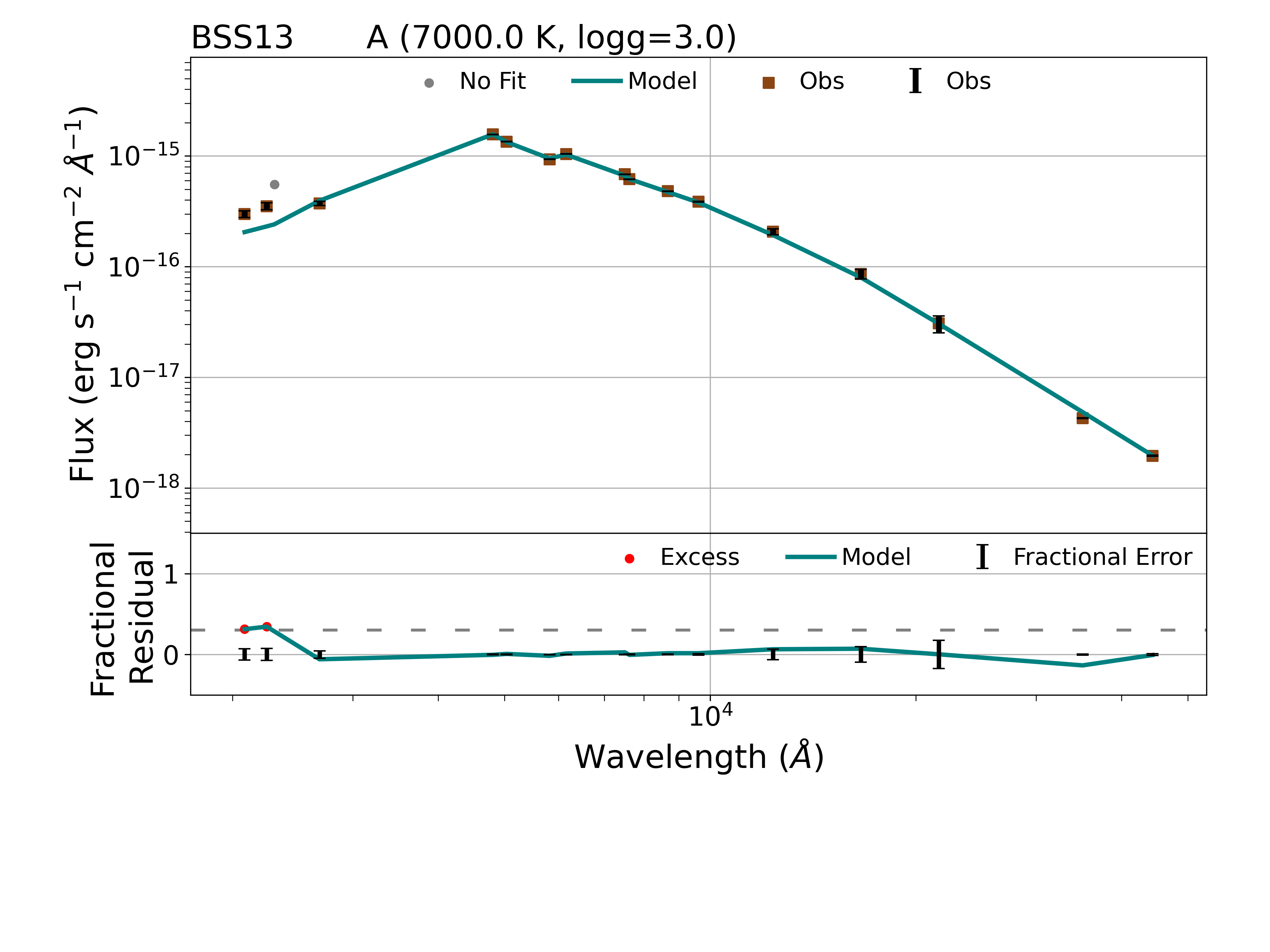}
		
	\end{minipage}%
	\centering
	\begin{minipage}{0.49\textwidth}
		\centering
		\includegraphics[width=\textwidth]{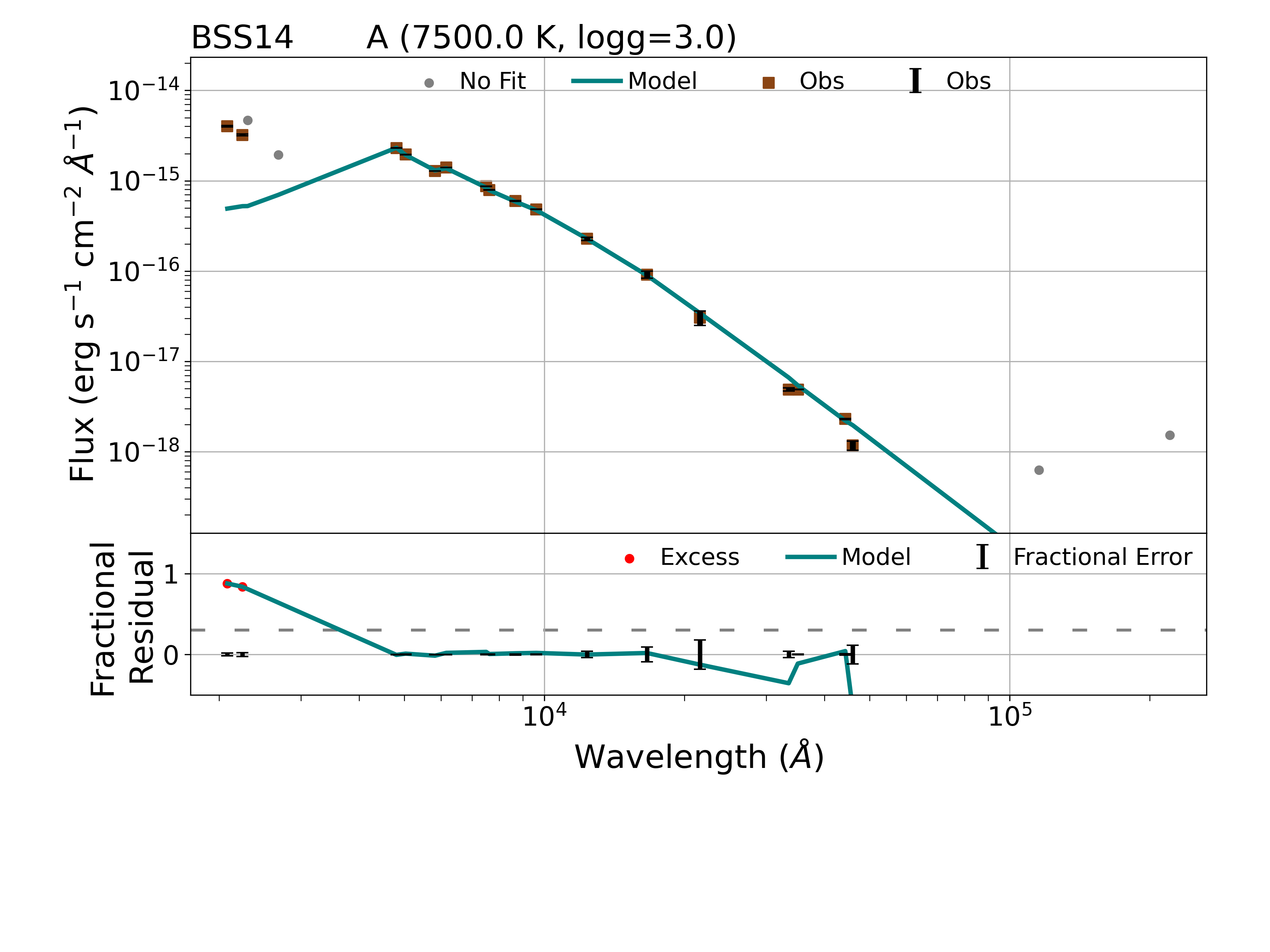}
		
	\end{minipage}%
		\hfill
	\begin{minipage}{0.49\textwidth}
		\centering
		\includegraphics[width=\textwidth]{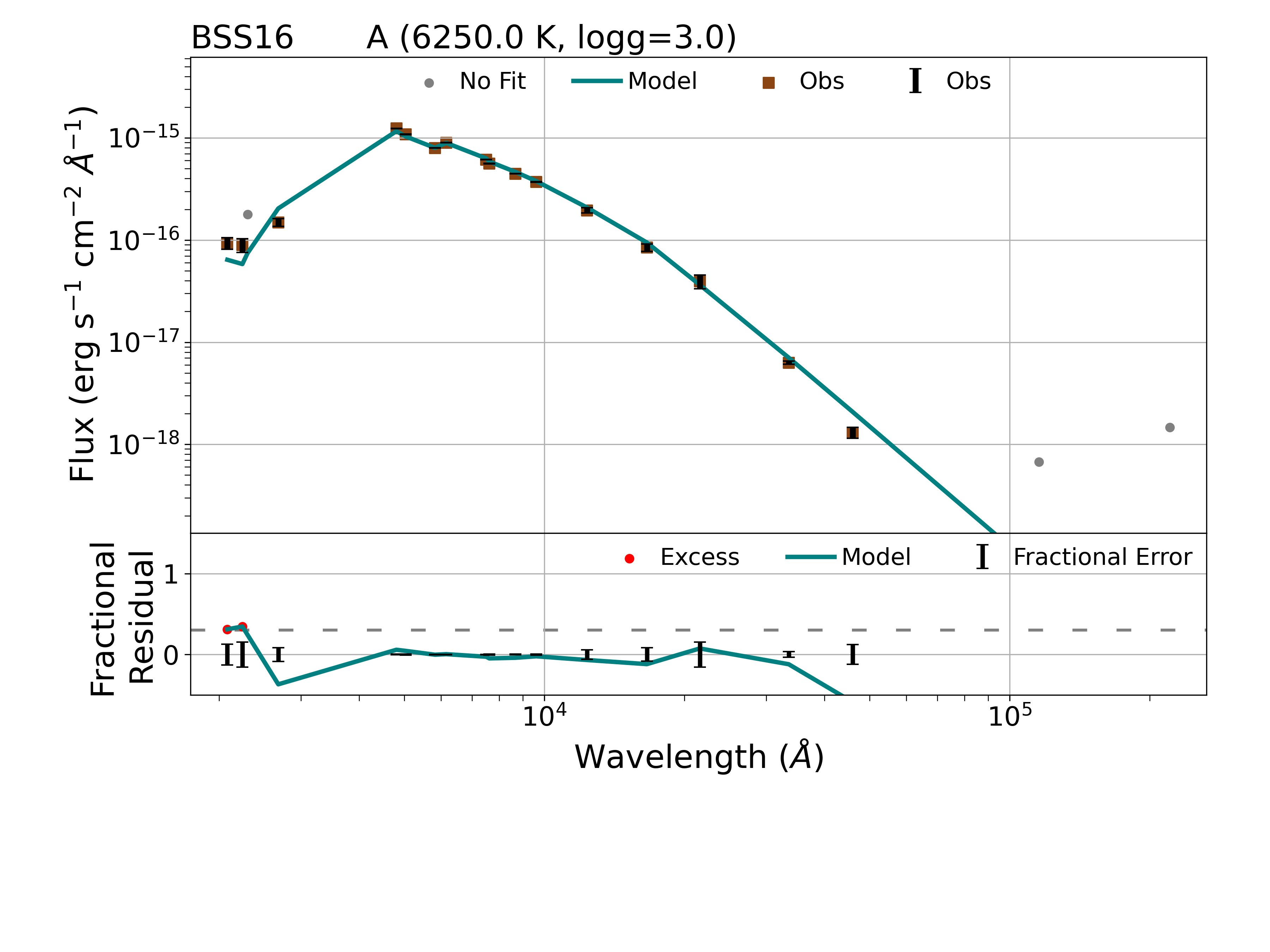}
		
	\end{minipage}%
	\caption{The single component SED fit of BSS having fractional residual greater than 0.3 in at least two UVOT filters. The symbols and curves mean the same as in Figure \ref{Figure3}.}
	\label{Figure5}
\end{figure*}

\begin{figure}
		\centering
		\includegraphics[width=0.5\textwidth]{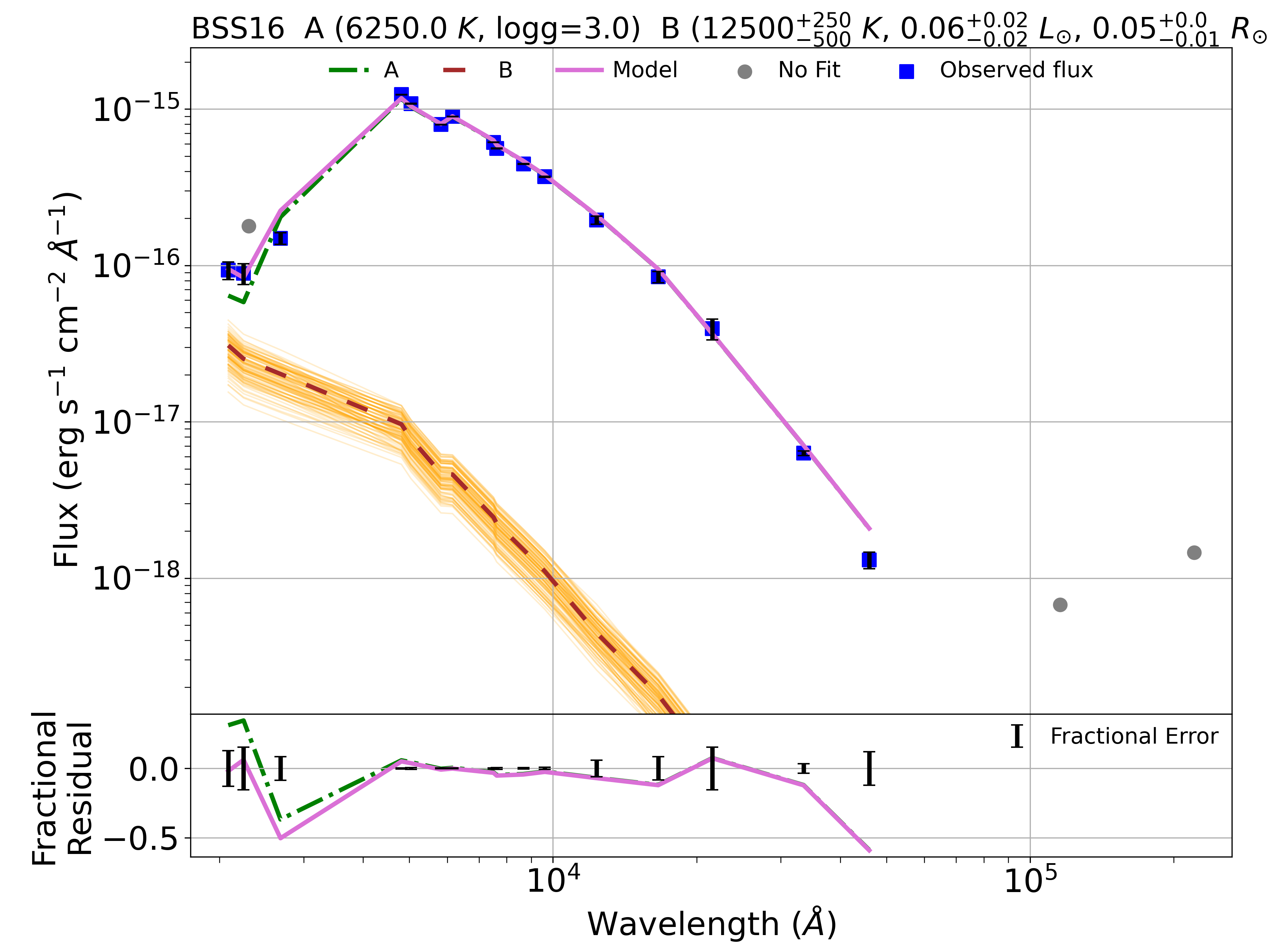}
    	\caption{The binary-component SED of BSS16. The top panel displays the double component SED of each object, with blue data points representing extinction corrected flux values (labeled as “Observed flux”) and black error bars showing flux errors. The green dashed line represents the cool component fit (labeled as “A”), while the brown dashed line represents the hot (B) component fit (labeled as “B”), with orange curves representing iteration residuals. The composite fit is depicted as a pink curve (labeled as “Model”), with grey data points denoting data points that were not included in the fits (labeled as “No fit”). The fractional residual for single (green) and composite (pink) fits is shown in the bottom panel. On the x-axis, black error bars represent fractional errors (labeled as “Fractional error”). The cool and hot component parameters derived from the SED, as well as the estimated errors, are listed at the top of the plots.}
    	\label{Figure6}
\end{figure}

Figure \ref{Figure3} shows the SEDs of 8 BSS and 2 YSS candidates successfully fitted with a single temperature. The residual of the fits is nearly zero, implying that the model fits well to the observed fluxes. Our criteria for considering sources for a double-component fit is a fractional residual greater than 0.3 in at least two UVOT filters. In the Table \ref{Table3}, we list the value of fractional residual for all the three UVOT filters for all the BSS and YSS candidates. Five BSS candidates show lower fractional residuals, less than 0.3, in all but one UVOT filter. We do not consider them for the double fit. The SEDs of these five BSS are shown in Figure \ref{Figure4}. As can be seen almost of all of them show an excess in the UVW2 filter which is the shortest wavelength among the three UV filters. There are three BSS candidates that exhibit fractional residuals exceeding 0.3 in at least two UVOT filters, thereby qualifying our criteria for a double-component SED fit. Such large UV excess could be due to the presence of a hot companion, chromospheric activity like hot spots in contact and semi-detached binaries, coronal emissions at very high temperatures, sub-luminous companions, magnetic activities of stars, flares on the stars, or active binaries. To understand the reason for the UV excess, we first searched for the X-ray counterparts of these sources. However, no X-ray data are available in \textit{Chandra} or \textit{XMM} for this cluster. In order to check the possibility of UV excess coming from an unresolved hot companion, we try to fit a double component SED using Koester \citep{Koester2010} and Kurucz \citep{Castelli1997} model. In case of two BSS, BSS13 and BSS14, we are unable to fit the observed UV excess using the two-component SEDs as the model suggests an upper limit of temperature, $\sim$80000 K, for the hot components. Hence, these two component fits are unreliable. The single-componenet SED fits of these three BSS with fractional residual greater than 0.3 are shown in Figure \ref{Figure5}. As can be seen, for BSS13, the excess is detected in two UVOT filters, UVW1 and UVM2, whereas for BSS14, the excess is detected in all the three UVOT filters, UVW1, UVM2 and UVW2, as well as the \textit{GALEX} NUV datapoint. The third BSS with fractional UV excess greater than 0.3 in at least two UVOT filters, BSS16, is successfully fitted with a double-component SED. We fit the hot component with the Koester model \citep{Koester2010} whereas the cool component is fitted with the Kurucz model  \citep{Castelli1997}. Based on the SED fit (see Figure \ref{Figure6}), we estimate that the the hot component has a temperature of 12500$\pm$125 K, a radius of 0.05$\pm$0.01 $R_\odot$, and a luminosity of 0.06$\pm$0.02 $L_\odot$. The parameters of all BSS, YSS, and their companion obtained from SED fits are listed in Table \ref{Table3}.

\pagebreak
\section{Discussion and Summary}\label{sec:4}
The availability of data from \textit{Swift}/UVOT for OC Berkeley 39 enabled us to study the exotic stars of this cluster in the NUV regime. To characterize BSS and YSS candidates of OC Berkeley 39, we utilized \textit{Swift}/UVOT data in three filters: UVW2, UVM2, and UVW1, along with the other multi-wavelength data. We first identified the cluster members using a machine-learning algorithm, ML-MOC, on \textit{Gaia} DR3 and obtained a total of 729 members, including 17 BSS and 4 YSS candidates. Out of the identified BSS and YSS candidates, we could study 16 BSS candidates and 2 YSS candidates since one BSS candidate and one YSS candidate lie outside the FOV of \textit{Swift}/UVOT, whereas one YSS candidate is detected only in one of the UVOT filters. To estimate the fundamental parameters (temperatures, radii, and luminosities) of BSS and YSS candidates, we constructed multi-wavelength SEDs using a virtual observatory tool, VOSA. Apart from obtaining the fundamental parameters of the exotic stars, our objective was to search for the presence of unresolved hot companions in these stars. Similar studies have been previously carried out to investigate exotic populations of OCs using HST by \cite{Geller, Gosnell}, and \cite{G}, \textit{Astrosat}/UVIT by \cite{Subramaniam2016, Sindhu2019, 2019ApJ...886...13J, Panthi2022, Vaidya7789, Anju6940, Anju2024} and \textit{Swift}/UVOT by \cite{Rao2022}.\\

\begin{figure*}[htbp]
	\centering
	\begin{minipage}{0.49\textwidth}
		\centering
		\includegraphics[width=\textwidth]{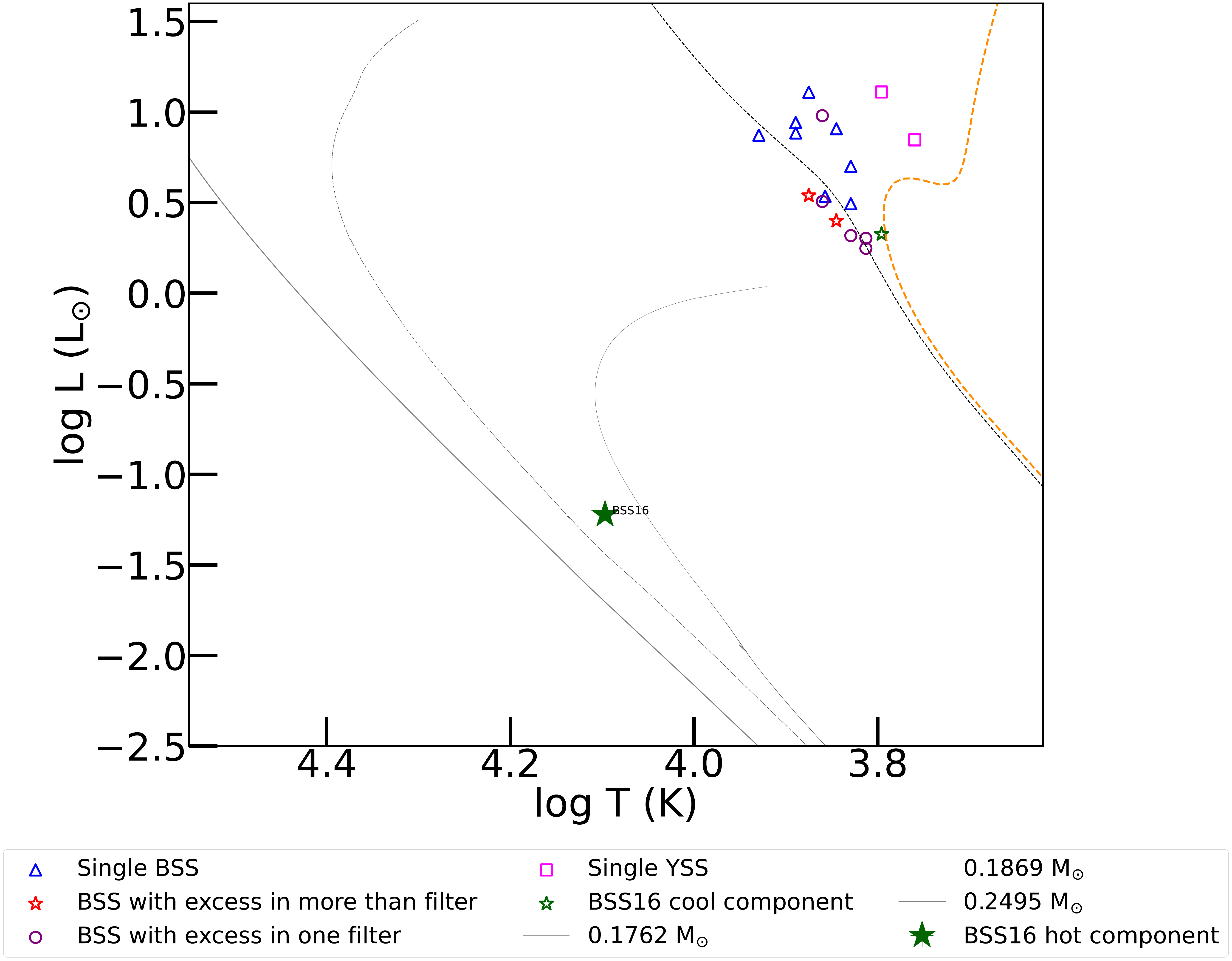}
		(a)
	\end{minipage}%
	\hfill
	\begin{minipage}{0.49\textwidth}
		\centering
		\vspace{-1cm}
		\includegraphics[width=\textwidth]{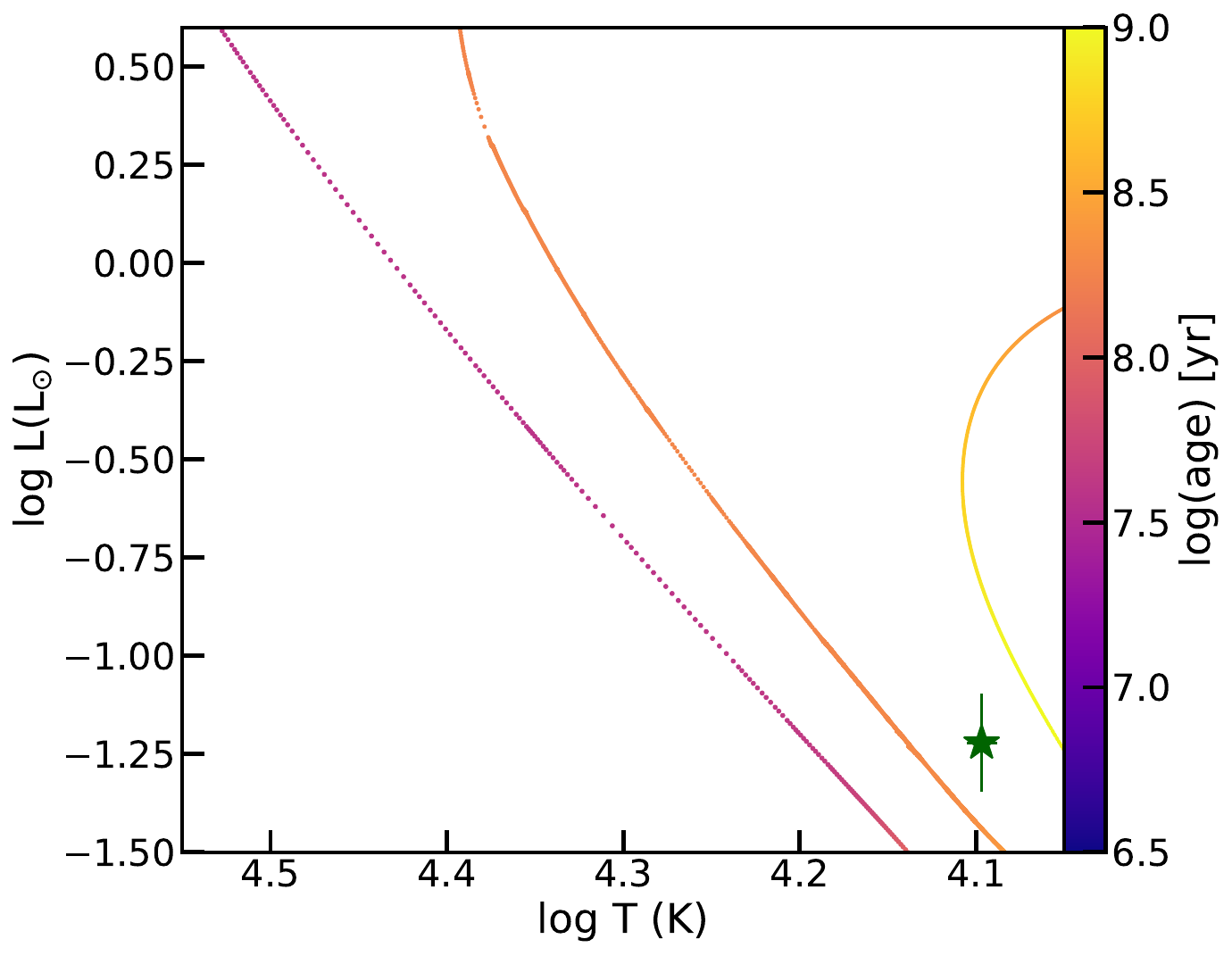}
		(b)
	\end{minipage}%
	\hfill
	\caption{(a) H-R diagram showing the single-component BSS as blue open triangles and YSS as pink open squares. BSS with excess in more than one filter are shown as stars whereas BSS with excess in one filter is shown as open circles. The cooler component of BSS16 (most probably a BL or just a binary star) is shown with a green open star, and the hot companion associated with it is shown in a green solid star. An orange dotted curve represents a 6 Gyr PARSEC isochrone, and the ZAMS is depicted as a black dashed curve. Various solid grey curves represent low mass (LM) and ELM WD cooling curves of different masses, sourced from \cite{Panie} and \cite{Althaus}.
		(b) The hot companion of BSS16 (most probably a BL or just a binary star) lies near the ELM WD cooling curves of 0.186 $M_\odot$ \citep{Panie}, indicating its approximate cooling age.}
	\label{Figure7}
\end{figure*}

Of the identified BSS and YSS candidates in \textit{Swift}/UVOT, 8 BSS and 2 YSS candidates fitted successfully with single-component SED. For these BSS, temperatures range from 6750 K to 8500 K, which is consistent given the age of the cluster, as also witnessed in a similar age OC M67 whose BSS have temperatures range from 6250 K -- 9000 K \citep{Sindhu2019}. The luminosities for these eight BSS vary from 3.12 $L_\odot$ to 12.89 $L_\odot$. For YSS, the temperatures range from 5750 K to 6250 K, and luminosities vary from 7.03 $L_\odot$ to 12.91 $L_\odot$. The lower temperature range of YSS suggests that YSS are cooler as compared to BSS, which is expected since YSS are believed to be evolved BSS \citep{Leiner2016}. On the H-R diagram, shown in Figure \ref{Figure7}, most of these eight BSS (open triangles), including BSS10 which is a $\delta$-Scuti variable \citep{Beata}, lie farther from the MSTO. Based on the extension of a single stellar isochrone of the cluster, their estimated mass gain ranges from 0.23 $M_\odot$ to 0.56 $M_\odot$ corresponding to a MSTO mass of 1.05 $M_\odot$. The absence of UV-excesses in their CMDs, as well as their locations on the H-R diagram (farther from the MSTO) suggest that they may have formed via mergers. However, it is also possible that some of them have formed through MT, but their companion WDs are currently cooler than $\sim$11000 K and therefore remain undetected.\\

Five BSS, including BSS6, which is likely an eclipsing binary \citep{Beata}, show small amount of UV excess; fractional residual up to 0.3 in UVW2 filter. Since an excess in a single UV filter does not provide us an opportunity to obtain a reliable fit, we fit them with single-componenet SEDs here. These excesses could be due to presence of hot spots or chromospheric activity, or even the presence of a hot companion. It is worth noting that the excess is visible in the UVW2 filter, which is the shortest wavelength among the three NUV filters. Therefore, it is likely that the observed excess in these SEDs increases in the far UV wavelengths. However, the far UV data for this cluster are not available at the moment. As per the single-component fits, their temperatures range from 6500 K to 7250 K. Four of the five BSS are located closer to the MSTO on the H-R diagram (open circles). Assuming the single stellar isochrone, they seem to have gained mass from 0.15 $M_\odot$  to 0.25 $M_\odot$ with the exception of BSS5, which shows a larger mass gain of 0.54 $M_\odot$. These smaller mass gains are consistent with MT by the Case-C formation mechanism \citep{Chen2008}. Therefore, the reason for the UV excess in a single filter is likely due to a hot companion, and these BSS may have formed via MT. However, for confirmation, far-UV data and/or spectroscopic observations are needed.\\

Objects that show moderate UV excess, fractional residual greater than 0.3 in at least two UV filters, are considered for double component SED fits. Three BSS meet this criterion, and we consider them for double component SED fits. The reasons for these UV excesses need to be looked into in further detail as the fractional residuals of 0.3 in at least two UV data points cannot be explained without either the presence of a WD companion or some other mechanisms, such as hot spots or chromospheric activities. The three BSS, BSS13, BSS14, and BSS16, have T$_\mathrm{eff}$ $\sim$6250 K -- 7500 K and L $\sim$2.12 $L_\odot$ -- 3.46 $L_\odot$ based on single-componenet SEDs. To examine the possible reasons for UV excess in these three BSS, we searched for X-ray data in the \textit{XMM} and \textit{Chandra} catalogues, but the data are not available for this cluster. We tried to fit a double SED component using the binary-component SED fitting code \citep{Jadhav2021} with Koester and Kurucz models. BSS16, having fractional residual greater than 0.3 in two of the UVOT filters, UVW2 and UVM2, fitted well with the Koester model corresponding to a hot component having a temperature, $ 12500^{+250}_{-500}$ K, a radius, 0.05$\pm$0.01 $R_\odot$, and a luminosity, 0.06$\pm$0.02 $L_\odot$. As can be seen in the H-R diagram (Figure \ref{Figure7}), the hot companion of BSS16 lies very close to the WD cooling curve of 0.186 $M_\odot$ \citep{Panie}, indicating it to be an ELM WD, assuming the UV excess is indeed due to the presence of a hot companion. Since ELM WDs cannot form through single star evolution within the age of the universe \citep{brown2011binary}, this suggests that the progenitor of this WD must have undergone significant mass loss during its evolutionary process, supporting the Case A/Case B MT formation mechanism of BSS16. From the cooling age of the ELM WD as shown in Figure \ref{Figure7}b, we can infer that the MT in this system ended recently, about 6.3 Myr ago. The primary component of BSS16 lies very close to and slightly below the MSTO in the H-R diagram, suggesting that this source could be called a BL.\\

For BSS13 and BSS14, the best-fitted double-component models reached the upper limit of the temperature for the hot component. Similar cases have been observed in other star clusters. In NGC 6940, one BL and one red clump star exhibited UV excess but could not be fitted with binary-component SEDs \citep{Anju6940}. Authors speculated that these two stars could be binary systems, and additional spectroscopic observations are needed to confirm. Alternatively, the UV excess may be due to other physical reasons such as hot spots, magnetic or chromospheric activities, or flares in a single or binary system. We need high-resolution spectroscopic observations to understand the reason for the UV excess in this system. The cool components of these two BSS are also lying near the MSTO suggesting a mass gain of less than 0.12 $M_\odot$, indicating their origin is MT in a binary if the UV excess is not originated due to chromospheric activity or hot excess.\\

To conclude, of the 16 BSS whose SEDs are presented here, we speculate that eight may have formed via mergers or MT with companion WDs cooler than $\sim$11000 K whereas the remaining eight may have formed via MT. Of these eight BSS that likely originated in MT, we are able to characterize the hot companion of a single BSS, BSS16. Based on the properties of this hot companion, it appears to be an ELM WD, suggesting that the BSS+ELM WD system formed as a result of Case-A/Case-B MT. In addition to that, one of these eight BSS includes BSS6, which has been previously identified as a likely eclipsing binary \citep{Beata} and, therefore, would have formed via the Case-A/Case-B formation mechanism. Using Swift/UVOT observations, \cite{Rao2022} found a single BSS out of 11 BSS to be a binary BSS with a hot companion based on the NUV excess. Since UVOT data are in the NUV filters, the detection of unresolved hot companions in \cite{Rao2022} and our work are smaller compared to those found in the FUV-based AstroSat/UVIT observations such as \cite{Vaidya7789,Anju6940}. This is expected since the excess due to hot WD companions is significant at wavelengths smaller than $\sim$1800 \AA. Therefore the number of BSS with unresolved hot companions in our work represents only a lower limit on the actual numbers of such BSS in this cluster.

\begin{table*}[!h]
	\centering
	\caption{For each BSS candidate and YSS candidate, \textit{Gaia} source ID in column 2, coordinates in columns 3 and 4, \textit{Swift}/UVOT UVW2, UVM2, and UVW1 fluxes and their associated errors are in columns 5, 6, and 7, respectively.}
	\label{Table2}
	\begin{tabular}{ccccccc}
		\hline
		\multicolumn{1}{c}{Name} & \multicolumn{1}{c}{GAIA DR3 source\textunderscore id} & \multicolumn{1}{c}{RA} &\multicolumn{1}{c}{DEC} & \multicolumn{3}{c}{UV flux (ergs s$^{-1}$ cm$^{-2}$ \AA$^{-1}$)} \\
		& &(deg)&(deg) & UVW2$\pm{\mathrm{err}}$ & UVM2$\pm{\mathrm{err}}$ & UVW1$\pm{\mathrm{err}}$ \\  
		\hline
		BSS1 & 3056682289893334912 & 116.61005 & $-$4.66228 & 2.40e-17$\pm$3.06e-18 & 2.09e-17$\pm$3.27e-18 & 8.13e-17$\pm$5.13e-18 \\
		BSS2 & 3056682083734884736 & 116.66626 & $-$4.66237 & 1.13e-16$\pm$6.21e-18 & 9.94e-17$\pm$6.92e-18 & 1.69e-16$\pm$6.72e-18 \\
		BSS3 & 3056682874008981888 & 116.68007 & $-$4.62433 & 5.39e-17$\pm$4.36e-18 & 6.23e-17$\pm$5.48e-18 & 1.02e-16$\pm$5.496e-18  \\
		BSS4 & 3056679060077913088 & 116.68629 & $-$4.68343 & 6.41e-16$\pm$1.46e-17 & 4.64e-16$\pm$1.47e-17 & 7.02e-16$\pm$1.32e-17  \\
		BSS5 & 3056678750840279296 & 116.69946  & $-$4.71175 & 3.71e-16$\pm$1.12e-17 & 3.24e-16$\pm$1.23e-17 & 5.53e-16$\pm$1.17e-17 \\
		BSS6 & 3056682702210133632 & 116.70154 & $-$4.62616 & 3.93e-17$\pm$3.81e-18 & 3.38e-17$\pm$4.07e-18 & 8.49e-17$\pm$5.19e-18  \\
		BSS7 & 3056679472394757504 & 116.70512 & $-$4.67469 & 1.54e-16$\pm$7.24e-18 & 1.42e-16$\pm$8.21e-18 & 3.56e-16$\pm$9.49e-18  \\
		BSS8 & 3056679231876595584 & 116.70588 & $-$4.68348 & 9.79e-17$\pm$5.81e-18 & 1.06e-16$\pm$7.09e-18 & 2.39e-16$\pm$7.99e-18 \\
		BSS9 & 3056679266236321024 & 116.71739 & $-$4.67424 & 3.09e-17$\pm$3.39e-18 & 2.04e-17$\pm$3.31e-18 & 8.58e-17$\pm$5.21e-18 \\
		BSS10 & 3056684214038601344 & 116.72419 & $-$4.60423 & 4.09e-16$\pm$1.17e-17 & 3.27e-16$\pm$1.23e-17 & 5.43e-16$\pm$1.15e-17  \\
		BSS11 & 3056679369315529728 & 116.73059 & $-$4.67562 & 4.90e-16$\pm$ 1.28e-17 & 4.42e-16$\pm$1.44e-17 & 7.45e-16$\pm$1.35e-17 \\
		BSS12 & 3056679609833686400 & 116.73173 & $-$4.65667 & 4.15e-16$\pm$1.18e-17 & 3.89e-16$\pm$1.35e-17 & 5.79e-16$\pm$1.19e-17 \\
		BSS13 & 3056684248398335616 & 116.73739 & $-$4.59998 & 7.30e-17$\pm$5.07e-18 & 8.54e-17$\pm$6.39e-18 & 1.33e-16$\pm$6.09e-18  \\
		BSS14 & 3056678441602599296 & 116.74244 & $-$4.69012 & 9.83e-16$\pm$1.81e-17 & 7.79e-16$\pm$1.91e-17 & 6.96e-16$\pm$1.31e-17\\
		BSS15 & 3056681121662156672 & 116.74742 & $-$4.63435 & 1.14e-16$\pm$6.26e-18 & 8.78e-17$\pm$6.47e-18 & 1.82e-16$\pm$7.03e-18 \\
		BSS16 & 3056678372883126528 & 116.75689 & $-$4.70461 & 2.28e-17$\pm$2.94e-18 & 2.16e-17$\pm$3.32e-18 & 5.34e-17$\pm$4.54e-18  \\
		YSS1 & 3056679094437661056 & 116.66802 & $-$4.68711 & 1.41e-17$\pm$2.45e-18 & 1.1e-17$\pm$2.46e-18 & 1.01e-16$\pm$5.69e-18\\
		YSS2 & 3056679644193433344 & 116.70590 & $-$4.65552 & .03E-17$\pm$4.99E-18 & 6.28e-17$\pm$5.52e-18 & 2.12e-16$\pm$7.46e-18\\
		
		\hline
	\end{tabular}
\end{table*}

\begin{table*}[!h]
	\centering
	\caption{The estimated parameters of the BSS candidates and YSS candidates using best-fitted SEDs. The dashed line in the column 2 represents the value of fractional residual less than 0.01.}
	\label{Table3}
	\begin{tabular}{cccccccccccc}
		\hline
		\multicolumn{1}{c}{Name} &\multicolumn{3}{c}{Fractional residual in UV}& \multicolumn{1}{c}{Component} &\multicolumn{1}{c}{log\,\textit{g}} &\multicolumn{1}{c}{$\chi_r^2$} &\multicolumn{1}{c}{$vgf_b$} & \multicolumn{1}{c}{Temperature}  & \multicolumn{1}{c}{Radius} & \multicolumn{1}{c}{Luminosity} \\
		& UVW2 & UVM2 & UVW1 & & & & & [K] & [$R_\odot$] &[$L_\odot$] \\  
		\hline
		BSS1 & 0.23 & 0.09 & 0.10 & single & 3.0 & 117.39 & 1.53 & 6500$\pm$125 & 1.05$\pm$0.11 & 1.77$\pm$0.35  \\
		BSS2 & 0.23 & 0.05 & -- & single & 3.0 & 67.93 & 1.01 & 7250$\pm$125 & 1.14$\pm$0.11 & 3.21$\pm$0.63 \\
		BSS3 & 0.01 & 0.02 & -- & single & 4.0 & 58.04 & 1.96 & 6750$\pm$125 & 1.29$\pm$ 0.13 & 3.12$\pm$ 0.62 \\
		BSS4 & 0.07 & -- & -- & single & 3.0 & 119.61 & 0.34 & 8500$\pm$125 & 1.26$\pm$0.12 & 7.46$\pm$0.15 \\
		BSS5 & 0.30 & 0.13 & -- & single & 3.0 & 33.55 & 0.21 & 7250$\pm$125 & 1.96$\pm$0.19 & 9.56$\pm$0.18  \\
		BSS6 & 0.22 & -- & -- & single & 3.0 & 78.16 & 1.71 & 6750$\pm$125 & 1.06$\pm$0.10  & 2.08$\pm$0.41   \\
		BSS7 & -- & -- & -- & single & 3.0 & 87.05 & 0.35 & 7000$\pm$125 & 1.93$\pm$0.19 & 8.09$\pm$0.16   \\
		BSS8 & 0.13 & 0.07 & -- & single & 4.0 & 122.92 & 0.77 & 6750$\pm$125 & 1.63$\pm$0.16 & 5.02$\pm$1.01  \\
		BSS9 & 0.32 & -- & 0.03 & single & 3.0  & 176.89 & 1.12 & 6500$\pm$125 & 1.12$\pm$0.11  &  2.01$\pm$0.39  \\
		BSS10 & 0.17 & -- & -- & single & 3.0 & 332.71 & 0.83 & 7750$\pm$125 & 1.54$\pm$0.15 & 7.67$\pm$1.51   \\
		BSS11 & 0.08 & -- & -- & single & 3.0 & 25.68 & 1.38  & 7500$\pm$125 & 2.13$\pm$0.21 & 12.89$\pm$2.53  \\
		BSS12 & 0.07 & -- & -- & single & 3.0 & 148.62 & 0.81 & 7750$\pm$125 & 1.64$\pm$0.16 & 8.75$\pm$1.71 \\
		BSS13 & 0.31 & 0.34 & -- & single & 3.0 & 76.93 & 0.29 & 7000$\pm$125 & 1.08$\pm$0.10 & 2.51$\pm$0.49  \\
		BSS14 & 0.87 & 0.83 & 0.64 & single & 3.0 & 58.76 & 3.51 & 7500$\pm$125 & 1.11$\pm$0.11 & 3.46$\pm$0.68   \\
		BSS15 & 0.18 & -- & -- & single &3.0 & 131.60 & 0.68 & 7200$\pm$125 & 1.17$\pm$0.12 & 3.43$\pm$0.68  \\
		BSS16 & 0.31 & 0.34 & --& single & 3.0 & 49.02 & 2.33 & 6250$\pm$125 & 1.24$\pm$0.12 &  2.12$\pm$0.42  \\
		           & -- & -- & -- & double & 3.0 & 45.28 & 4.14 & $12500^{+250}_{-500}$ & 0.05$\pm$0.01 &  0.06$\pm$0.02  \\
		YSS1 & -- & 0.05 & -- & single & 4.0  & 45.19 & 0.31 & 5750$\pm$125 & 2.66$\pm$0.21 & 7.03$\pm$1.12  \\
		YSS2 & 0.06 & 0.03 & -- & single & 3.5  & 215.68 & 0.26 & 6250$\pm$125 & 3.03$\pm$0.24 & 12.91$\pm$2.07  \\
		\hline
	\end{tabular}
\end{table*}

\section{Acknowledgements}
The authors would like to thank Manan Agarwal for developing the ML-MOC algorithm, using which the cluster membership has been determined for this work. This work has made use of the third data release from the European Space Agency (ESA) mission \textit {Gaia} (\url{https://www.cosmos.esa.int/gaia}), \textit{Gaia} DR3 \citep{GaiaDR3}, processed by the \textit {Gaia} Data Processing and Analysis Consortium (DPAC, \url{https://www.cosmos.esa.int/web/gaia/dpac/consortium}). This research made use of ASTROPY, a PYTHON package for astronomy \citep{Astropy}, NUMPY \cite{harris2020array}, MATPLOTLIB \cite{hunter2007matplotlib}. This research also made use of the Astrophysics Data System (ADS) governed by NASA (\url{https://ui.adsabs.harvard.edu}).

\section{Data Availability}
The data underlying this article are publicly available at \url{https://gea.esac.esa.int/archive}. The derived data generated in this research will be shared upon reasonable request to the corresponding author.


\bibliography{sample631}{}
\bibliographystyle{aasjournal}



\end{document}